# Continuous atomic displacements and lattice distortions during martensitic transformations in fcc-bcc-hcp systems

Cyril Cayron

Laboratory of ThermoMechanical Metallurgy (LMTM), PX Group Chair, Ecole Polytechnique Fédérale de Lausanne (EPFL), Rue de la Maladière 71b, 2000 Neuchâtel, Switzerland. cyril.cayron@epfl.ch

**Abstract:** This work generalizes our previous works on fcc→bcc martensitic transformations to the larger family of transformations in the fcc-bcc-hcp system and to fcc→fcc mechanical twinning. The analytical expressions of the atomic displacements and lattice distortions are calculated directly from the orientation relationships without any adjustment of free parameter; the unique assumption is that the atoms are hard-spheres that can't interpenetrate themselves. The habit planes are predicted on the simple criterion that they are unrotated by the distortion, and the results are compared to experimental observations published in literature. It is shown that shuffle is required for transformations implying the hcp phase because the hcp primitive Bravais lattice contains two atoms, instead of one for the fcc and bcc phases. A simple encoding of the lattice distortions and shuffles permits to attribute a groupoid structure to the transformations in the fcc-hcp-bcc system. The analytical expressions of the intermediate states are given and could be used to calculate activation energies. The martensitic distortion occurs in one-step, without shearing, and its accommodation generates orientation gradients in the parent phase, independently of its glide modes. The concept of reversibility is detailed on the basis of crystallographic and morphological arguments. The possibility to apply this approach to diffusion-limited martensitic transformations is discussed.

## 1. Introduction

Shearing is the corner stone of crystallographic calculations in metallurgy, for twinning [1]-[6] and more generally for martensitic transformations [7]-[12] in face-centered-cubic (fcc), body-centered-cubic (bcc) and hexagonal close-packed phases. Shear matrices take the form [7]

$$\mathbf{S} = \mathbf{I} + \mathbf{d}.{}^t\mathbf{p} = \mathbf{I} + \mathbf{d}\otimes\mathbf{p} \qquad (1)$$

where $\mathbf{I}$ is the identity matrix, $\mathbf{d}$ is the (column) shear direction and ${}^t\mathbf{p}$ is the transpose of the normal of the shear plane (it is a row vector). $\mathbf{d}$ and $\mathbf{p}$ are vectors of the direct and reciprocal spaces respectively. The displacement vector can be written $\mathbf{d} = \mathbf{s} + \delta$ with $\mathbf{s}$ belonging to the plane $\mathbf{p}$, i.e. $\mathbf{s}.\mathbf{p} = \mathbf{0}$, and $\delta$ parallel to $\mathbf{p}$. When $\mathbf{d}$ and $\mathbf{p}$ are perpendicular, $\delta = 0$, the strain $\mathbf{s}$ is a simple shear, as it is the case for twinning. The phenomenological theories of martensitic crystallography (PTMC) make an extensive use of shear matrices [7]-[12]. Since a simple shear does not change the volume of the crystal, transformations between phases of different densities imply shear that is not simple; the volume change is given by the term δ and by the determinant of Bain distortion matrix in the other side of the PTMC equation.

It is undoubtable that a shear stress induces twinning or martensitic transformation in metastable alloys, but actually, the lattice deformation accompanying these mechanisms can't be a simple shear deformation. Shear deformation conveniently describes the lattice distortion from the initial state to the final transformed state, but if the size of the atoms in the lattice is taken into account in the intermediate states, it is clear that a shear would make the atoms interpenetrate (Fig. 1), which is energetically unattainable for metals or metallic alloys. In recent papers [13][14], we have shown that it is possible to establish a simple crystallographic approach of fcc→bcc martensitic transformations without requiring to any shear matrix. In paper [13], we focused our efforts on the base of a Pitch distortion, and the idea was generalized in the paper [14] to Kurdjumov-Sachs orientation relationship (OR). The atoms are considered as hard-spheres and the model combines the four sequential matrices of PTMC into a unique "angular distortive" matrix. It was shown that this matrix has two unrotated planes: the low-index {111} and the high-index {225} planes, in good agreement with the experimentally determined habit planes (HP). The aim of the present paper is to further generalize the approach to the wide family of phase transformations between fcc, bcc and hcp phases, so important in metallurgy because of their applications in Fe, Ti, Zr, Co and many other alloys. The case of fcc→fcc twinning is also treated.

The first part of the paper is mainly mathematical; the chapters 3 to 6 are dedicated to fcc→fcc twinning and fcc→hcp , bcc→fcc and bcc→hcp transformations, respectively. Each chapter follows the same scheme: a) the matrix of complete transformation is calculated from the OR, then b) the intermediate states of the distorted lattice and atomic displacements are determined; shuffle is calculated when required, and c) the habit planes are numerically deduced and briefly compared to experimental results given in literature. The style is voluntarily repetitive in order to point out the similarities of the transformations. A mathematical unification of the transformations in the fcc-hcp-bcc system is proposed in chapter 7. The main ideas of the model and their consequences in term of qualitative understanding and quantitative predictability are discussed in chapter 8.

## 2. Notations and elementary formulae

Let us call $\mathbf{B}_0^\gamma$= ($\mathbf{a}_0^\gamma$, $\mathbf{b}_0^\gamma$, $\mathbf{c}_0^\gamma$) with $\mathbf{a}_0^\gamma$ = $[100]_\gamma$, $\mathbf{b}_0^\gamma$ = $[010]_\gamma$, $\mathbf{c}_0^\gamma$ = $[001]_\gamma$, the reference basis of the γ phase. The distortion matrix can be calculated by finding a primitive lattice of the parent phase and following how this lattice is transformed. Let us call $\mathbf{B}_p^\gamma$ this starting primitive basis, and $\mathbf{B}_p^{\gamma'}$ its distorted image. The latter is a primitive basis of the product phase α. The two bases are expressed by the matrices $\mathbf{B}_p^\gamma = [\mathbf{B}_0^\gamma \to \mathbf{B}_p^\gamma]$ and $\mathbf{B}_p^{\gamma'} = [\mathbf{B}_0^\gamma \to \mathbf{B}_p^{\gamma'}]$, respectively. The distortion matrix expressed in the initial primitive basis is $\mathbf{D}_p^{\gamma\to\alpha} = [\mathbf{B}_p^\gamma \to \mathbf{B}_p^{\gamma'}] = [\mathbf{B}_p^\gamma \to \mathbf{B}_0^\gamma][\mathbf{B}_0^\gamma \to \mathbf{B}_p^{\gamma'}] = (\mathbf{B}_p^\gamma)^{-1}\mathbf{B}_p^{\gamma'}$. In the reference basis of the parent phase it is

$$\mathbf{D}_0^{\gamma\to\alpha} = \left[\mathbf{B}_0^{\gamma} \to \mathbf{B}_p^{\gamma}\right]\mathbf{D}_p^{\gamma\to\alpha}\left[\mathbf{B}_p^{\gamma} \to \mathbf{B}_0^{\gamma}\right] = \mathbf{B}_p^{\gamma'}\left(\mathbf{B}_p^{\gamma}\right)^{-1} \quad (2)$$

This formula can be used to calculate the distortion matrix when the images of the vectors of the primitive lattice are known in the reference basis of the parent phase. If these images are known only in the reference lattice of the product phase, one can use the matrix of change of coordinates $C_0^{\gamma\to\alpha}$ which gives the coordinates of the vectors of the reference basis $\mathbf{B}_0^{\alpha}$ of the $\alpha$ crystal in the reference basis $\mathbf{B}_0^{\gamma}$ of the $\gamma$ crystal:

$$\mathbf{C}_0^{\gamma\to\alpha} = \left[\mathbf{B}_0^{\gamma} \to \mathbf{B}_0^{\alpha}\right] \quad (3)$$

The correspondence matrix can be determined directly from the orientation relationship, as explained for example in ref. [13]. Therefore, if the images of vectors of the primitive lattice are known only in the coordinate system of the final phase and are expressed by the three column vectors forming the matrix $\mathbf{B}_p^{\alpha}$, it is possible to convert them into the initial system by writing

$$\mathbf{B}_p^{\gamma'} = \mathbf{C}_0^{\gamma\to\alpha}\mathbf{B}_p^{\alpha}$$

and then applying equation (3) to determine the distortion matrix. Equation (3) will be used to determine the distortion matrix of the complete transformation. For all the intermediate states, the calculations are similar but the size of the atoms should be taken into account to avoid their interpenetration.

In the paper, the fcc phase will be noted $\gamma$, the bcc phase $\alpha$, and the hcp phase $\varepsilon$, as it is usually done for steel (but unfortunately this notation is different for brass). We will use the three-index notation for the hcp phase. As already assumed in ref. [13] and [14] for steels, the metal atoms are considered as hard spheres of same diameter in all the phases, which implies that:

$$2a_{\varepsilon} = \sqrt{2}\, a_{\gamma} = \sqrt{3}\, a_{\alpha} \quad (4)$$

The ORs that will be used for the fcc-hcp-bcc transformations are those that respect the parallelism of the close-packed directions, i.e. Kurdjumov-Sachs (KS) [15] for fcc→bcc (actually discovered by Young [16] in iron meteorites few years before Kurdjumov and Sachs), Burgers [17] for bcc→hcp, and Shoji-Nishiyama (SN) [8] for fcc→hcp. These ORs are

- KS: $[110]_{\gamma} = [111]_{\alpha}$ and $(\bar{1}11)_{\gamma} // (\bar{1}10)_{\alpha}$ (5)
- Burgers: $[111]_{\alpha} = [100]_{\varepsilon}$ and $(\bar{1}10)_{\alpha} // (001)_{\varepsilon}$
- SN : $[110]_{\gamma} = [100]_{\varepsilon}$ and $(\bar{1}11)_{\gamma} // (001)_{\varepsilon}$

It is possible to define an orthonormal basis $\mathbf{B}_c$ common to the fcc, bcc and hcp crystals in KS-Burgers-SN ORs. This basis is given in each reference basis by

- $\left[\mathbf{B}_0^{\gamma} \to \mathbf{B}_c\right] = \frac{1}{a_\gamma}\begin{bmatrix} 1/\sqrt{2} & 1/\sqrt{6} & 1/\sqrt{3} \\ 1/\sqrt{2} & -1/\sqrt{6} & -1/\sqrt{3} \\ 0 & 2/\sqrt{6} & -1/\sqrt{3} \end{bmatrix}$

- $\left[\mathbf{B}_0^{\alpha} \rightarrow \mathbf{B}_c\right] = \frac{1}{a_\alpha}\begin{bmatrix} 1/\sqrt{3} & 1/\sqrt{6} & -1/\sqrt{2} \\ 1/\sqrt{3} & 1/\sqrt{6} & 1/\sqrt{2} \\ 1/\sqrt{3} & -2/\sqrt{6} & 0 \end{bmatrix}$
- $\left[\mathbf{B}_0^{\varepsilon} \rightarrow \mathbf{B}_c\right] = \frac{1}{a_\varepsilon}\begin{bmatrix} 1 & 1/\sqrt{3} & 0 \\ 0 & 2/\sqrt{3} & 0 \\ 0 & 0 & \sqrt{3/8} \end{bmatrix}$ (6)

These matrices associated to theoretical ratio of the lattice parameters given in equation (4) allow the calculation of the correspondence matrices between the crystals of two phases. For example, $C_0^{\gamma\rightarrow\alpha}$ = $\left[\mathbf{B}_0^{\gamma} \rightarrow \mathbf{B}_0^{\alpha}\right] = \left[\mathbf{B}_0^{\gamma} \rightarrow \mathbf{B}_c\right]\left[\mathbf{B}_c \rightarrow \mathbf{B}_0^{\alpha}\right] = \left[\mathbf{B}_0^{\gamma} \rightarrow \mathbf{B}_c\right]\left[\mathbf{B}_0^{\alpha} \rightarrow \mathbf{B}_c\right]^{-1}$. There values are reported in Table 1. It can be checked that the circular products is equal to the identity matrix. For example,

$$\mathbf{C}_0^{\gamma\rightarrow\alpha}\mathbf{C}_0^{\alpha\rightarrow\varepsilon}\mathbf{C}_0^{\varepsilon\rightarrow\gamma} = \mathbf{I} \tag{7}$$

where **I** is the identity matrix.

Some comparisons will be realized between the calculated and experimental habit planes reported in literature. The equivalent symmetries of the parent and daughter phases may make this comparison difficult. Therefore, in order to keep the coherency of notation in the whole paper, the reported OR and corresponding HP will be written according to the choice of equation (5), if possible without ambiguity.

## 3. FCC→FCC mechanical twinning

Twinning in fcc materials can appear under different circumstances. Annealing twins are created at high temperatures during grain recrystallization and growth. They are formed along straight {111} planes “by accident” during the reorientation of the new grains because there is nearly no energy difference between the new crystal and its Σ3 twins. Generally, all the four Σ3 variants are formed, and the process can repeat itself during the recrystallization which generates $\Sigma 3^n$ twins [18][19]. At high and room temperature, narrow twinned bands, called microtwins, can also appear under strains. It is usually assumed that these microtwins are formed by regular creation and synchronized displacements of Shockley partial dislocations on the close-packed {111} planes transforming the ABCABC stacking into an ACBACB twinned stacking. This has been extensively confirmed by in-situ Transmission Electron Microscopy (TEM) observations, but the exact mechanism at the origin of the twinning dislocations is not yet fully understood. It is believed to occur via a pole mechanism (initially proposed by Cottrell and Bilby for twinning in bcc crystals [20]): a screw dislocation spirals around a pole and allows a layer-by-layer shearing on adjacent parallel {111} planes [21]. To our knowledge, no spiraling dislocations in microtwins could have been evidenced, but recent TEM observations in Cu nano-alloys actually show that grain boundaries could be the sources of the Shockley partial dislocations [22]. Microtwinning is an important mode of deformation in the Twinning Induced Plasticity (TWIP) and Transformation Induced Plasticity (TRIP) steels [23][24].

At very low temperatures, large mechanical twins can appear massively under strains by bursts producing load instability. These macrotwins propagates at speeds close to the speed of sound; which gives rise to audible clicks (twinning "cry"), as clearly shown by Blewitt *et al* [25] in Cu-8%Al single crystal deformed at 4.2K (their study is also reported in [6]). It is difficult to understand how a pole mechanism could be in agreement with such high speeds. Actually, it is probable that the mechanical twins result from a collective motion of the atoms as for fcc-bcc martensitic transformation. In that assumption, the atoms would move cooperatively in one step. As mentioned in the introduction, classical approaches of the lattice deformations yielded by macrotwinning and martensitic transformation are based on shears [1]-[5]; however, the crystallographic link between the two mechanisms is not clearly established. The aim of this section is to show that it is possible to define fcc→fcc mechanical macrotwinning without using shear matrices, but by using angular distortive matrices, exactly as it was done for fcc→bcc martenstic transformations [14]. Although fcc→fcc macrotwinning is not very important in metallurgy because it occurs at very low temperatures, the comparison is worth being explained; moreover, the results will be used to help the calculations in the other sections.

### 3.1. Matrix of complete lattice distortion

As for fcc→bcc transformation in ref. [13] and [14], we choose $(\bar{1}11)_\gamma$ as unrotated plane. For twinning, this plane is actually invariant. The reference frame and the positions of the atoms in the initial fcc state are shown in Fig. 2 and Fig. 3b. The triangle POK is unchanged by twinning (β=60°), contrarily to what happens in the fcc→bcc transformation. The atom in M, initially such that **PM** = $[100]_\gamma$, moves and climbs between the two atoms in O and K and, after twinning, in its final position, M is located such that the tetrahedron POKM is regular. The lattice distortion of complete transformation can be determined by considering that the vectors **x** = **PO** =½ $[110]_\gamma$ and **y** = **PK** =½ $[101]_\gamma$ are invariant, and that the vector **z** = **PM** = 1/3 $[211]_\gamma$ + 1/3 $[1\bar{1}\bar{1}]_\gamma$ = $[100]_\gamma$ is transformed into the vector **PM'** = 1/6 $[211]_\gamma$ +1/3 $[1\bar{1}\bar{1}]_\gamma$ = 1/6 $[4\bar{1}\bar{1}]_\gamma$. This means that

$$\mathbf{B}_p^\gamma = \begin{bmatrix} 1/2 & 1/2 & 1 \\ 1/2 & 0 & 0 \\ 0 & 1/2 & 0 \end{bmatrix} \text{ and } \mathbf{B}_p^{\gamma'} = \begin{bmatrix} 1/2 & 1/2 & 4/6 \\ 1/2 & 0 & -1/6 \\ 0 & 1/2 & -1/6 \end{bmatrix}$$

, and thus, by using formula (2),

$$\mathbf{D}_0^{\gamma\to\gamma} = \mathbf{B}_p^{\gamma'}\left(\mathbf{B}_p^\gamma\right)^{-1} = \frac{1}{6}\begin{bmatrix} 4 & 2 & 2 \\ -1 & 7 & 1 \\ -1 & 1 & 7 \end{bmatrix} \qquad (8)$$

### 3.2. Matrix of continuous lattice distortion

The matrix (8) gives the complete twinning distortion of the fcc lattice. It is possible to determine all the intermediate states of the transformation by considering that the atoms are hard-spheres that "roll" on each other. Let us call η the angle (**JH**, **JM**), i.e. the angle between the $(\bar{1}11)_\gamma$ and $(111)_\gamma$ planes. For fcc twinning the deformation occurs such that $\mathrm{JP} = \mathrm{JM} = \sqrt{3}/_2\, d$ where *d* is the atom diameter, thus $JP = JM = \sqrt{6}/_4\, a_\gamma$. Since the triangle PJM remains isosceles during the distortion, the angle η is such that η = 2γ (here the angle γ should not be confused with the fcc γ phase), as represented in Fig. 2b and Fig. 3b.

During the displacement of point M, the angle η varies from η = arcos(1/3) = 70.5° to η = - arcos(1/3) = 180°-70.5°. Twinning is obtained by the same displacement of the atom in M as for the fcc→bcc transformation [14] except that the angle β remains fixed at 60°, which means that the atom in M has to climb between the two atoms in O and K such that the distance OK remains constant and equal to *d*, whereas in the fcc→bcc martensite transformation, this distance increases by the opening of the angle (**PO**, **PK**) from 60° to 70.5°. Therefore, equation 21 of ref. [14] of fcc→bcc transformation can be modified in order to get its equivalent for fcc→fcc twinning. It becomes

$$\left[\mathbf{B}_s \rightarrow \mathbf{B}_p(\eta)\right] = \begin{bmatrix} 1 & \cos(\beta) & \cos(\gamma)\|z\|\cos(\beta/2) \\ 0 & \sin(\beta) & \cos(\gamma)\|z\|\sin(\beta/2) \\ 0 & 0 & \|z\|\sin(\eta/2) \end{bmatrix} \tag{9}$$

with β = 60°, η = 2γ and $\|\boldsymbol{z}\|$ the norm of the vector **z** = $[100]_\gamma$ = **PM**. Equation (9) gives the evolution of the values of the basis $\mathbf{B}_p$ of the primitive lattice formed by the vectors **PO**, **PK** and **PM** during the transformation in reference to a fixed orthonormal basis $\mathbf{B}_s$ formed from the vectors **PO**, **PK** and **PM** in the initial fcc crystal. The correspondence matrix from $\mathbf{B}_0^\gamma$ to $\mathbf{B}_p$ = (**PO**, **PK**, **PM**) in the initial fcc reference basis is:

$$\left[\mathbf{B}_0^\gamma \rightarrow \mathbf{B}_p(\eta = 70.5°)\right] = \begin{bmatrix} 1/\sqrt{2} & 1/\sqrt{2} & 1 \\ 1/\sqrt{2} & 0 & 0 \\ 0 & 1/\sqrt{2} & 0 \end{bmatrix} \tag{10}$$

One should consider the following points in equation (9): the distortion matrix of twinning is expressed in the $\mathbf{B}_s$ basis, β = 60°, the part $\cos(\gamma)\|\boldsymbol{z}\|$ is the projection of **PM** on the line **PJ** and is equal to $\mathrm{PJ}(1+\cos(\gamma))$, and $\sin(\gamma)\|\boldsymbol{z}\| = PJ\sin(\eta)$, as illustrated in Fig. 2c. Therefore, equation (9) can also be expressed by:

$$\left[\mathbf{B}_s \rightarrow \mathbf{B}_p(\eta)\right] = \begin{bmatrix} 1 & 1/2 & \frac{3}{4\sqrt{2}}(1+\cos(\eta)) \\ 0 & \sqrt{3}/2 & \frac{\sqrt{3}}{4\sqrt{2}}(1+\cos(\eta)) \\ 0 & 0 & \frac{\sqrt{3}}{2\sqrt{2}}\sin(\eta) \end{bmatrix} \tag{11}$$

The lattice distortion matrix in the reference basis $\mathbf{B}_p$ is

$$\mathbf{D}_p^{\gamma\rightarrow\gamma}(\eta) = \left[\mathbf{B}_p(70.5°) \rightarrow \mathbf{B}_p(\eta)\right] = \left[\mathbf{B}_p(70.5°) \rightarrow \mathbf{B}_s\right]\left[\mathbf{B}_s \rightarrow \mathbf{B}_p(\eta)\right]$$

with [$\mathbf{B}_p(70.5°) \rightarrow \mathbf{B}_s$] = [$\mathbf{B}_s \rightarrow \mathbf{B}_p(70.5°)$ ]$^{-1}$ also given by equation (11).

It is now possible to calculate the twinning lattice distortion in the reference basis $\mathbf{B}_0^\gamma$ by using formula (2):

$$\mathbf{D}_0^{\gamma\to\gamma}(\eta) = [\mathbf{B}_0^\gamma \rightarrow \mathbf{B}_p]\, \mathbf{D}_p^{\gamma\to\gamma}(\eta) [\mathbf{B}_0^\gamma \rightarrow \mathbf{B}_p]^{-1} \tag{12}$$

$$= \frac{1}{4}\begin{bmatrix} 2+2Y+\sqrt{2}\sqrt{1-Y^2} & 2-2Y-\sqrt{2}\sqrt{1-Y^2} & 2-2Y-\sqrt{2}\sqrt{1-Y^2} \\ 1+Y-\sqrt{2}\sqrt{1-Y^2} & 3-Y+\sqrt{2}\sqrt{1-Y^2} & -1-Y+\sqrt{2}\sqrt{1-Y^2} \\ 1+Y-\sqrt{2}\sqrt{1-Y^2} & -1-Y+\sqrt{2}\sqrt{1-Y^2} & 3-Y+\sqrt{2}\sqrt{1-Y^2} \end{bmatrix}$$

with Y = cos(η).

It can be checked that for the initial state, since η = 70.5°, Y = cos(η) = 1/3, $\mathbf{D}_0^{\gamma\to\gamma}(\eta = 70.5°)$ is the identity matrix. The complete transformation matrix is obtained for η = 180°-70.5°, Y = cos(η) = -1/3, which leads to the matrix given in equation (8). This is the same expression as it could be found by a simple shear of amplitude *s* = 1/√2 on a $(\bar{1}11)_\gamma$ plane on the -$[211]_\gamma$ direction [10]. The interesting point here is that all the continuous intermediate states can be calculated from equation (12), whereas a continuous simple shear is unrealistic due to the interpenetration of the atoms, i.e. the atom M displaced by shearing would collide with the atoms in O and K.

All the atoms M of the crystal move during the fcc→fcc twinning transformation exactly as the lattice ones:

$$\mathbf{PM'} = \mathbf{D}_0^{\gamma\to\gamma}(\eta)\, \mathbf{PM} \tag{13}$$

Therefore, as for fcc→bcc transformation, mechanical twinning does not require shuffle.

### 3.3. Habit plane

The matrix $\mathbf{D}_0^{\gamma\to\gamma}$ gives the images in the initial fcc basis $\mathbf{B}_0^\gamma$ of the direction $\mathbf{u}_0^\gamma$ by twinning:

$$\mathbf{u}'^\gamma_0 = \mathbf{D}_0^{\gamma\to\gamma}\mathbf{u}_0^\gamma \tag{14}$$

The inverse of its transpose gives the images of the plane $\mathbf{g}_0^\gamma$:

$$\mathbf{g}'^\gamma_0 = \left(\mathbf{D}_0^{\gamma\to\gamma}\right)^* \mathbf{g}_0^\gamma \text{ , with} \tag{15}$$

$$\left(\mathbf{D}_0^{\gamma\to\gamma}\right)^* = {}^T\left(\mathbf{D}_0^{\gamma\to\gamma}\right)^{-1} = \frac{1}{6}\begin{bmatrix} 8 & 1 & 1 \\ -2 & 5 & -1 \\ -2 & -1 & 5 \end{bmatrix} \tag{16}$$

It can be checked that this matrix has an invariant (reciprocal) vector; it is the $(\bar{1}11)_\gamma$ plane which therefore appears the natural habit plane for mechanical twinning, as expected.

## 4. FCC→ HCP transformation

Martensitic fcc→hcp transformations occur in cobalt alloys at temperatures which depends strongly on the alloy composition (the $M_s$ temperature of pure cobalt is 422°C) [26]-[28], in some Fe-Cr-Ni stainless steels quenched at low temperatures [29][30], and in some Mn-rich steels with shape memory properties [31]. The crystallographic explanations of the transformation given by metallurgists rely on arguments very similar to those used for fcc microtwinning: regular arrays of Shockley partial dislocations gliding on the $\{111\}_\gamma$ planes change the ABCABC stacking order of the fcc phase into the ABABAB order of the hcp phase. However, to our knowledge, there is no consensus on the exact sequence of creation, dissociation and glide of dislocations at the origin of the nucleation and growth of the hcp phase [26][27]. A completely different model was proposed by the physicists P. Toledano *et al*. [32]. In their model, both the fcc and hcp structures in cobalt result from a "reconstructive" ordering mechanism of a disordered polytypic structure (for physicists the term "reconstructive" is not synonymous of "diffusive", see section 8.1).

To our point of view, none of the explanations are totally satisfactory. Models based on coordinated creation and displacements of partial dislocations are not compatible with the high speeds of martensitic transformations. The model based on the disordered intermediate lattice does not take into account the atom size to explain how the atoms could move during the fcc→hcp reordering process. Let us consider now that the fcc→hcp transformation occurs quickly in one-step as for fcc→fcc macrotwinning, i.e. without coordinated motions of partial dislocations or the without the existence an intermediate latent lattice.

### 4.1. Matrix of complete lattice distortion

In order to calculate the lattice distortion matrix, an intermediate primitive basis $\mathbf{B}_p^\gamma$ = (**x**, **y**, **z**)$_\gamma$ should be found. Clearly, since the vectors **PO** = ½ $[110]_\gamma$ and **PK** = ½ $[101]_\gamma$ are invariant, they can be chosen for the **x** and **y** axes, respectively, as for twinning. The vector **PM** cannot be taken as **z** axis of intermediate basis $\mathbf{B}_p^\gamma$ because the (**x**, **y**, **z**)$_\gamma$ and the distortion would be exactly the same as for twinning. It will be shown that the displacement of the atom in M is a shuffle. Instead, the vector **PN** = ½ $[2\bar{1}\bar{1}]_\gamma$ appears appropriate (Fig. 3c), as it will be proved in the following section. We also point out here that contrarily to the primitive basis $\mathbf{B}_p^\gamma$ used for fcc→bcc transformations, the basis $\mathbf{B}_p^\gamma$ used in this section contains not one, but two atoms.

This lattice distortion matrix can be determined similarly as for the fcc→fcc twinning transformation; however, now, instead of choosing the atom M located on the $(\bar{1}11)_\gamma$ plane at the level on the *l = 1* (Fig. 3c), we choose the atom in N located at the level *l = 2*, i.e. such that **PN** = ½ $[2\bar{1}\bar{1}]_\gamma$. The vectors **PO** = ½ $[110]_\gamma$ and **PK** = ½ $[101]_\gamma$ of invariant plane $(\bar{1}11)_\gamma$ are invariant, and the vector **PN** = 1/6 $[211]_\gamma$ +2/3 $[1\bar{1}\bar{1}]_\gamma$ = ½ $[2\bar{1}\bar{1}]_\gamma$ is transformed into the vector **PN'** = 2/3 $[1\bar{1}\bar{1}]_\gamma$. This means that

$\mathbf{B}_p^\gamma = \begin{bmatrix} 1/2 & 1/2 & 1 \\ 1/2 & 0 & -1/2 \\ 0 & 1/2 & -1/2 \end{bmatrix}$ and $\mathbf{B}_p^{\gamma'} = \begin{bmatrix} 1/2 & 1/2 & 2/3 \\ 1/2 & 0 & -2/3 \\ 0 & 1/2 & -2/3 \end{bmatrix}$, and thus, by using formula (2),

$$\mathbf{D}_0^{\gamma\to\varepsilon} = \mathbf{B}_p^{\gamma'}\left(\mathbf{B}_p^{\gamma}\right)^{-1} = \frac{1}{12}\begin{bmatrix} 10 & 2 & 2 \\ -1 & 13 & 1 \\ -1 & 1 & 13 \end{bmatrix} \quad (17)$$

### 4.2. Matrix of continuous lattice distortion

The matrix (17) gives the complete distortion of the fcc lattice. Here again, it is possible to determine all the intermediate states of the transformation by considering that the atoms are hard-spheres that "roll" on each other. During the fcc→hcp transformation the points P, O, K, I and J are fixed (Fig. 2a and Fig. 3c). The atom N has to jump above the two atoms located on the ($\bar{1}11$) plane below. There are two possibilities for the atom located in M: it can jump above the atoms located in O and K, following the same trajectory as for the atom N, but one ($\bar{1}11$) plane below N, or it can remain in the same position. If the atom in M moves, its trajectory can be deduced from the calculations performed for twinning in the previous section. In the $\mathbf{B}_0^{\gamma}$ basis, the vector **JM** = ¼ $[2\bar{1}\bar{1}]_\gamma$ is transformed exactly as for twinning, i.e. into the vector **JM'** = $\mathbf{D}_0^{\gamma\to\gamma}(\eta)$ **JM** with $\mathbf{D}_0^{\gamma\to\gamma}(\eta)$ the matrix given in equation (12). The vector **IN** follow the same change as the vector **JM**, and thus **IN'** = $\mathbf{D}_0^{\gamma\to\gamma}(\eta)$ **IN**. The vector **PN** is thus changed into **PN'** = **PI** + **IN'** = **PI** + $\mathbf{D}_0^{\gamma\to\gamma}(\eta)$ **IN** .Therefore, the primitive basis $\mathbf{B}_p$ formed by the vectors **x** = **PO** = ½ $[110]_\gamma$ , **y** = **PK** = ½ $[101]_\gamma$ and **z** = **PN = PI + IN** takes the form of a matrix (**x**, **y**, **z**) that can be expressed directly in the $\mathbf{B}_0^{\gamma}$ basis by:

$$\left[\mathbf{B}_0^{\gamma} \to \mathbf{B}_p(\eta)\right] = \begin{bmatrix} 1/2 & 1/2 & 1/2 + \frac{1}{4}(2Y + \sqrt{2}\sqrt{1-Y^2}) \\ 1/2 & 0 & -1/4 + \frac{1}{4}(Y - \sqrt{2}\sqrt{1-Y^2}) \\ 0 & 1/2 & -1/4 + \frac{1}{4}(Y - \sqrt{2}\sqrt{1-Y^2}) \end{bmatrix} \quad (18)$$

with Y = cos(η).

The continuous lattice distortion matrix deduced from formula (2) is

$$\mathbf{D}_0^{\gamma\to\varepsilon}(\eta) = \left[\mathbf{B}_0^{\gamma} \to \mathbf{B}_p(70.5°)\right] \left[\mathbf{B}_p(\eta) \to \mathbf{B}_0^{\gamma}\right]^{-1}$$
$$= \frac{1}{8}\begin{bmatrix} 6+2Y+\sqrt{2}\sqrt{1-Y^2} & 2-2Y-\sqrt{2}\sqrt{1-Y^2} & 2-2Y-\sqrt{2}\sqrt{1-Y^2} \\ 1+Y-\sqrt{2}\sqrt{1-Y^2} & 7-Y+\sqrt{2}\sqrt{1-Y^2} & -1-Y+\sqrt{2}\sqrt{1-Y^2} \\ 1+Y-\sqrt{2}\sqrt{1-Y^2} & -1-Y+\sqrt{2}\sqrt{1-Y^2} & 7-Y+\sqrt{2}\sqrt{1-Y^2} \end{bmatrix} \quad (19)$$

It can be checked that for the initial state, since η = 70.5°, Y = cos(η) = 1/3, $\mathbf{D}_0^{\gamma\to\varepsilon}(\eta = 70.5°)$ is the identity matrix. The complete transformation matrix is obtained for η = 180°-70.5°, Y = cos(η) = -1/3; which leads to the matrix (17).

### 4.3. Schuffle

All the atoms of type N = (*u*, *v*, *w*) located as N in the plane $(\bar{1}11)$ in even layers, i.e. such that *l = (-u+v+w)* is even, have a trajectory that directly follows the lattice distortion:

$$\mathbf{PN'} = \mathbf{D}_0^{\gamma\to\varepsilon}(\eta)\,\mathbf{PN} \tag{20}$$

All the other atoms of type M = (*u*, *v*, *w*) located as M in the plane $(\bar{1}11)$ in odd layers, i.e. such that *l = (-u+v+w)* is odd, have a trajectory that does not follow the lattice distortion, but that can be deduced of it. There are two equivalent shuffles of the M atoms that can be determined by considering the trajectories of M in its local unit cell. Either M does not move, or it moves as it would do for a twinning distortion. The origin P of the unit cell in which the atom M is located is deduced from M by the translation vector **t** = $[010]_\gamma$. Therefore, the trajectories of the atoms M that do not move in their unit cells obey the equation:

Shuffle S0: $$\mathbf{PM'} = \mathbf{D}_0^{\gamma\to\varepsilon}(\eta)\,(\mathbf{PM}-\mathbf{t}) + \mathbf{t} \tag{21}$$

And the trajectories of the atoms M that move by a local twinning displacement in their unit cells, as shown in Fig. 3c, obey the equation:

Shuffle S2: $$\mathbf{PM'} = \mathbf{D}_0^{\gamma\to\varepsilon}(\eta)\,(\mathbf{PM}-\mathbf{t}) + \mathbf{D}_0^{\gamma\to\gamma}(\eta)\,\mathbf{t}, \text{ with } \mathbf{t} = [010]_\gamma \tag{22}$$

These two trajectories (S0 and S2) do not have the same expression (20) as for the other atoms of the lattice; they are shuffles. The need of shuffle comes from the fact that the primitive unit cell of the hcp structure contains two atoms; which also explains why we could not choose **PM** in the primitive basis $\mathbf{B}_p$ and thus justifies our choice of basis at the beginning of section.

### 4.4. Habit plane

The matrix $\mathbf{D}_0^{\gamma\to\varepsilon}$ gives the images of the direction $\mathbf{u}_0^\gamma$ in the initial fcc basis $\mathbf{B}_0^\gamma$ by the fcc→hcp distortion.

The inverse of its transpose gives the image of the plane $\mathbf{g}_0^\gamma$.

$$\mathbf{g'}_0^\gamma = \left(\mathbf{D}_0^{\gamma\to\varepsilon}\right)^* \mathbf{g}_0^\gamma \text{ , with} \tag{23}$$

$$\left(\mathbf{D}_0^{\gamma\to\varepsilon}\right)^* = {}^T\left(\mathbf{D}_0^{\gamma\to\varepsilon}\right)^{-1} = \frac{1}{12}\begin{bmatrix} 14 & 1 & 1 \\ -2 & 11 & -1 \\ -2 & -1 & 11 \end{bmatrix} \tag{24}$$

It can be checked that this matrix has an invariant (reciprocal) vector which is $(\bar{1}11)_\gamma$, which therefore appears as the natural habit plane for fcc→hcp transformation.

The continuous analytical expressions of the angular distortive matrices of fcc→bcc transformations -equation 31 of ref. [14]-, of fcc→fcc macrotwinning -equation (12)-, and of fcc→hcp transformations -equation (19) with equation (21) for the shuffle-, have been introduced into a computer program written in VPython that allows the representation of crystals in three dimensions. Simulation "movies" of the distortion of a fcc cube constituted by 8x8x8 unit cells transformed into bcc, fcc-twinned and hcp structures are given in Supplementary Materials. The initial, intermediate and final states are represented in blue, yellow and red colors, respectively, for fcc→bcc, fcc→fcc and fcc→hcp transformations, in Fig. 4a,b and c, respectively.

## 5. BCC→FCC transformation

Bcc→fcc transformations are less frequent. They can be encountered in Fe-Cr-Ni duplex steels in which δ ferrite decomposes into lath and spearhead isolated austenite [33] or in Widmanstätten austenite at the δ grain boundaries [34]. Bcc→fcc transformations are also widely studied in Cu-Zn brass and other Cu-Al, Cu-Sn alloys. The bcc phase is ordered itself during cooling and is transformed into a B2 structure. The B2 phase undergoes a martensitic transformation by cooling below room temperature to form a monoclinic 9R structure which can be seen as a slightly distorted form of a polytype of the fcc (3R) phase [35]-[37]. In all these cases, the parent bcc and daughter fcc phases are in KS OR. Most often in literature the fcc or 9R daughter phase are created by thermal decomposition during an homogenisation at high temperature in the bcc domain (800-900°C) and by a thermal treatment at medium temperature (300-500°C) [38]-[40]. The bcc→fcc or bcc→9R transformation also occurs under strain and is at the origin of the shape memory effects in these alloys [35][36][40]. To our knowledge, there is no alloy in which pure martensitic fcc phase is formed under cooling. However, let us consider the ideal case where the atoms would move collectively from a bcc to a fcc structure. It will be discussed in section 8.5 whether or not such an approach can be applied to diffusion-limited transformations.

### 5.1. Matrix of complete lattice distortion

The distortion matrix can be determined by considering that the vectors **x** = $\mathbf{PO}_\alpha$ =½ $[111]_\alpha$ and **y** = $\mathbf{PK}_\alpha$ =½ $[11\bar{1}]_\alpha$ of the bcc phase become by lattice distortion the vectors $\mathbf{PO}_\gamma$ =½ $[110]_\gamma$ and $\mathbf{PK}_\gamma$ =½ $[101]_\gamma$ of the fcc phase (Fig. 5 and Fig. 6a), and that the vector $\mathbf{PM}_\alpha$ = $[010]_\alpha$ is transformed into the vector **z** = $\mathbf{PM}_\gamma$ = $[100]_\gamma$. It implies that the primitive basis

$\mathbf{B}_p^\alpha = \begin{bmatrix} 1/2 & 1/2 & 0 \\ 1/2 & 1/2 & 1 \\ 1/2 & -1/2 & 0 \end{bmatrix}$ has for image in $\mathbf{B}_0^\gamma$ the basis of the product phase, the basis

$$\mathbf{B}_{p/\gamma_0}^{\alpha'} = \begin{bmatrix} 1/2 & 1/2 & 1 \\ 1/2 & 0 & 0 \\ 0 & 1/2 & 0 \end{bmatrix}.$$

This image can be calculated in the reference basis of the parent crystal $\mathbf{B}_0^\alpha$ by using the correspondence matrix $\mathbf{C}_0^{\alpha\to\gamma}$ given in Table 1. It follows that $\mathbf{B}_p^{\alpha'} = \mathbf{C}_0^{\alpha\to\gamma}\mathbf{B}_{p/\gamma_0}^{\alpha'}$ and then $\mathbf{D}_0^{\alpha\to\gamma} = \mathbf{C}_0^{\alpha\to\gamma}\mathbf{B}_{p/\gamma_0}^{\alpha'}\left(\mathbf{B}_p^\alpha\right)^{-1}$, which becomes after calculations

$$\mathbf{D}_0^{\alpha\to\gamma} = \begin{bmatrix} \frac{3}{4}+\frac{\sqrt{6}}{24} & \frac{\sqrt{6}}{12} & \frac{1}{4}-\frac{\sqrt{6}}{8} \\ -\frac{1}{4}+\frac{\sqrt{6}}{24} & 1+\frac{\sqrt{6}}{12} & \frac{1}{4}-\frac{\sqrt{6}}{8} \\ \frac{1}{4}-\frac{\sqrt{6}}{12} & \frac{1}{2}-\frac{\sqrt{6}}{6} & \frac{1}{4}+\frac{\sqrt{6}}{4} \end{bmatrix} \tag{25}$$

Another way to find this result is to use the complete distortion matrix of the fcc→bcc transformation $\mathbf{D}_0^{\gamma\to\alpha}$ reported in equation 32 of ref. [14], and notice that

$$\mathbf{D}_0^{\alpha\to\gamma} = \mathbf{C}_0^{\alpha\to\gamma}\left(\mathbf{D}_0^{\gamma\to\alpha}\right)^{-1}\mathbf{C}_0^{\gamma\to\alpha} \tag{26}$$

### 5.2. Matrix of continuous lattice distortion

The lattice distortion matrix of the bcc→fcc transformation can be calculated from the fcc→bcc one, i.e. $\mathbf{D}_0^{\gamma\to\alpha}(\beta)$ given in our previous paper [14], but not by a simple inversion. This matrix is a function of the distortion angle β which varies from 60° to 70.5° from the fcc to the bcc structure. $\mathbf{D}_0^{\gamma\to\alpha}(60°)$ is the identity matrix; it lets the $\mathbf{B}_0^\gamma$ unchanged. $\mathbf{D}_0^{\gamma\to\alpha}(70.5°)$ is the complete transformation matrix; it transforms $\mathbf{B}_0^\gamma$ into $\mathbf{B}_0^{\gamma'}$ a basis of the bcc structure expressed into $\mathbf{B}_0^\gamma$; it is important to notice that $\mathbf{B}_0^{\gamma'}$ is not $\mathbf{B}_0^\alpha$. Actually $\mathbf{B}_0^{\gamma'}$ is constituted by the vectors $[100]_\alpha$ and $[110]_\alpha$ and $[\bar{1}10]_\alpha$. Therefore, the bcc→fcc distortion matrix $\mathbf{D}_0^{\alpha\to\gamma}(\beta)$ is not simply the inverse of the fcc→bcc distortion matrix $\mathbf{D}_0^{\gamma\to\alpha}(\beta)$; one has also to calculate it in the reference basis of $\mathbf{B}_0^\alpha$. For that aim, let us split the fcc→bcc path into two successive paths: the first one from $\mathbf{B}_0^\gamma$ to $\mathbf{B}_0^\gamma$ (β) = $\mathbf{D}_0^{\gamma\to\alpha}(\beta)\,\mathbf{B}_0^\gamma$, and the next one from $\mathbf{B}_0^\gamma$ (β) to $\mathbf{B}_0^{\gamma'}$ = $\mathbf{B}_0^\gamma$ (70.5°) = $\mathbf{D}_0^{\gamma\to\alpha}$ (70.5°) $\mathbf{B}_0^\gamma$. This decomposition expressed in $\mathbf{B}_0^\gamma$ takes the form:

$$\mathbf{D}_0^{\gamma\to\alpha} = \left[\mathbf{B}_0^\gamma \to \mathbf{B}_0^\gamma(\beta)\right]\left[\mathbf{B}_0^\gamma(\beta) \to \mathbf{B}_0^\gamma(70.5°)\right] = \mathbf{D}_0^{\gamma\to\alpha}(\beta)\,(\mathbf{D}_0^{\alpha\to\gamma}(\beta))^{-1} \tag{27}$$

Indeed, the last path is the inverse of the path from the basis $\mathbf{B}_0^{\gamma'}$, which is a bcc basis, to the intermediate basis, both expressed in $\mathbf{B}_0^\gamma$; i.e. it is the inverse of the bcc→fcc distortion matrix expressed $\mathbf{B}_0^\gamma$. The term "0" is not sufficient to avoid any confusion in the reference basis of the matrices $\mathbf{B}_0^\gamma$ or $\mathbf{B}_0^\alpha$; thus, we write $\mathbf{D}_{/\gamma 0}^{\alpha\to\gamma}(\beta)$ in order to specify that the bcc→fcc distortion matrix $\mathbf{D}_0^{\alpha\to\gamma}(\beta)$ in equation (27) is written in $\mathbf{B}_0^\gamma$. From equation (27), it follows that

$$\mathbf{D}^{\alpha\to\gamma}_{/\gamma 0}(\beta) = \left(\mathbf{D}^{\gamma\to\alpha}_{0}\right)^{-1} \mathbf{D}^{\gamma\to\alpha}_{/\gamma 0}(\beta) \tag{28}$$

The two terms at the right of this equation are known from ref.[14]. They are the inverse of the matrix of complete transformation and the matrix of transformation at intermediate state given by the angle β. Eventually, the bcc→fcc transformation matrix $\mathbf{D}^{\alpha\to\gamma}_{/\gamma 0}(\beta)$ can be expressed in the reference basis $\mathbf{B}^{\alpha}_{0}$ by using the correspondence matrix $\mathbf{C}^{\alpha\to\gamma}_{0} = \left[\mathbf{B}^{\alpha}_{0} \to \mathbf{B}^{\gamma}_{0}\right]$ given in Table 1. It becomes:

$$\mathbf{D}^{\alpha\to\gamma}_{0}(\beta) = \mathbf{D}^{\alpha\to\gamma}_{/\alpha 0}(\beta) = \mathbf{C}^{\alpha\to\gamma}_{0}\ \mathbf{D}^{\alpha\to\gamma}_{/\gamma 0}(\beta)\ \mathbf{C}^{\gamma\to\alpha}_{0} \tag{29}$$

The symbolic calculations have been performed with Mathematica. It leads to the components $d^{\alpha\to\gamma}_{ij}(\beta)$ of the matrix $\mathbf{D}^{\alpha\to\gamma}_{0}(\beta)$ expressed as function of X = cos(β):

(30)

$$d^{\alpha\to\gamma}_{11} = \frac{1}{72}(48 - 4\sqrt{6} + 9(-1 + 3\sqrt{6})\sqrt{X}\sqrt{\frac{1-X}{1+X}} + (11\sqrt{3} + 12\sqrt{2})\sqrt{1-X^2} + X(-48 + 8\sqrt{6} - (27\sqrt{3} + 9\sqrt{2})\sqrt{\frac{1-X}{1+X}}))$$

$$d^{\alpha\to\gamma}_{12} = \frac{1}{72}(-4\sqrt{6} + 9(1 - 3\sqrt{6})\sqrt{X}\sqrt{\frac{1-X}{1+X}} + (-7\sqrt{3} + 6\sqrt{2})\sqrt{1-X^2} + X(48 + 8\sqrt{6} + (27\sqrt{3} + 9\sqrt{2})\sqrt{\frac{1-X}{1+X}}))$$

$$d^{\alpha\to\gamma}_{13} = \frac{1}{36}(12 + 4\sqrt{6} - 8\sqrt{6}X - (2\sqrt{3} + 9\sqrt{2})\sqrt{1-X^2})$$

$$d^{\alpha\to\gamma}_{21} = \frac{1}{72}(48 - 4\sqrt{6} - 9(1 + 5\sqrt{6})\sqrt{X}\sqrt{\frac{1-X}{1+X}} + (11\sqrt{3} + 12\sqrt{2})\sqrt{1-X^2} + X(-48 + 8\sqrt{6} - (27\sqrt{3} + 9\sqrt{2})\sqrt{\frac{1-X}{1+X}}))$$

$$d^{\alpha\to\gamma}_{22} = \frac{1}{72}(-4\sqrt{6} + 9(1 + 5\sqrt{6})\sqrt{X}\sqrt{\frac{1-X}{1+X}} + (-7\sqrt{3} + 6\sqrt{2})\sqrt{1-X^2} + X(48 + 8\sqrt{6} + (27\sqrt{3} + 9\sqrt{2})\sqrt{\frac{1-X}{1+X}}))$$

$$d^{\alpha\to\gamma}_{23} = \frac{1}{36}(12 + 4\sqrt{6} - 8\sqrt{6}X - (2\sqrt{3} + 9\sqrt{2})\sqrt{1-X^2})$$

$$d^{\alpha\to\gamma}_{31} = \frac{1}{72}(48 - 4\sqrt{6} - 9(-2 + \sqrt{6})\sqrt{X}\sqrt{\frac{1-X}{1+X}} - (16\sqrt{3} + 15\sqrt{2})\sqrt{1-X^2} + X(-48 + 8\sqrt{6} + (54\sqrt{3} - 9\sqrt{2})\sqrt{\frac{1-X}{1+X}}))$$

$$d^{\alpha\to\gamma}_{32} = \frac{1}{72}(-4\sqrt{6} + 9(-2 + \sqrt{6})\sqrt{X}\sqrt{\frac{1-X}{1+X}} + (20\sqrt{3} - 21\sqrt{2})\sqrt{1-X^2} + X(48 + 8\sqrt{6} - (54\sqrt{3} - 9\sqrt{2})\sqrt{\frac{1-X}{1+X}}))$$

$$d^{\alpha\to\gamma}_{33} = \frac{1}{18}(6 + 2\sqrt{6} - 4\sqrt{6}X - (\sqrt{3} - 9\sqrt{2})\sqrt{1-X^2})$$

This matrix is a function of the distortion angle β which varies from β = 70.5° (X= 1/3) to β =60° (X=1/2) from the bcc to the fcc structure. One can check that $\mathbf{D}^{\alpha\to\gamma}_{0}(70.5°)$ is the identity matrix. The complete transformation is given by $\mathbf{D}^{\alpha\to\gamma}_{0} = \mathbf{D}^{\alpha\to\gamma}_{0}(60°)$, which is the matrix given in equation (25).

All the atoms M of the crystal are displaced during the bcc→fcc transformation exactly as those of the lattice:

$$\mathbf{PM'} = \mathbf{D}_0^{\alpha\to\gamma}(\beta)\,\mathbf{PM} \tag{31}$$

The bcc→fcc transformation does not require shuffle.

### 5.3. Prediction of the habit plane

The complete transformation matrix $\mathbf{D}_0^{\alpha\to\gamma}$ in equation (25) gives the images of the directions by the bcc→fcc distortion in the reference basis $\mathbf{B}_0^{\alpha}$. The images of the planes are given by the inverse of its transpose:

$$\left(\mathbf{D}_0^{\alpha\to\gamma}\right)^* = {}^{T}\left(\mathbf{D}_0^{\alpha\to\gamma}\right)^{-1} = \begin{bmatrix} 1+\frac{\sqrt{6}}{18} & \frac{\sqrt{6}}{18} & \frac{1}{3}-\frac{\sqrt{6}}{6} \\ -\frac{1}{3}+\frac{\sqrt{6}}{18} & \frac{2}{3}+\frac{\sqrt{6}}{18} & \frac{1}{3}-\frac{\sqrt{6}}{6} \\ \frac{1}{3}-\frac{\sqrt{6}}{9} & \frac{1}{3}-\frac{\sqrt{6}}{9} & \frac{1}{3}+\frac{\sqrt{6}}{3} \end{bmatrix} \tag{32}$$

The two matrices $\mathbf{D}_0^{\alpha\to\gamma}$ and $\left(\mathbf{D}_0^{\alpha\to\gamma}\right)^*$ give the images of the directions and planes in the initial basis $\mathbf{B}_0^{\alpha}$. These images can be calculated in the final base $\mathbf{B}_0^{\gamma}$ by using the correspondence matrix $\mathbf{C}_0^{\gamma\to\alpha}$ given in Table 1. The results obtained on the low-index directions and planes are given in Table 2. This table can be compared with the one given for the fcc→bcc transformation (Table 1 of ref. [14]). It can be noticed that the results are identical if the directions and planes are exchanged.

As in ref. [14], it is also possible to determine the habit plane as the plane **g** unrotated by the distortion:

$$\left\|\Delta \mathbf{g}_{\perp}\right\| = 0 \tag{33}$$

The rotation amplitude of a plane **g** (normalized reciprocal vector) is given in Fig. 7 as function of the spherical coordinate angles θ and ϕ of **g**. As for fcc→bcc transformation [14], four solutions are found, that can be grouped in two pairs due to the fact that the solutions at θ > π/2 are equivalent to those obtained at θ ≤ π/2 because of the centrosymmetric equivalence of lines (θ, ϕ) ≡ (π-θ, ϕ+π). Thus, there are two non-equivalent solutions, one at $(\theta = 1.571, \phi = 2.356)$ which corresponds to the $(\bar{1}10)_\alpha$ plane, and the other one at $(\theta = 1.222, \phi = 2.616)$ which corresponds to the plane $hkl = (-0.813, 0.471, 0.341)_\alpha$. This plane is at 1.3° of the rational plane $(\bar{5}32)_\alpha$. This plane is actually exactly the image of the habit plane predicted for the fcc→bcc transformation. Indeed, we recall here that it was a found that the unrotated plane for fcc→bcc was a plane at 0.5° of the $(\bar{2}25)_\gamma$ plane and that this plane was transformed into a plane close to $(\bar{7}43)_\alpha$ by the fcc→bcc distortion matrix. The small difference between the $(\bar{5}32)_\alpha$ and the $(\bar{7}43)_\alpha$ planes come from the rationalization of the vectors. We have checked that when the numerical values are taken into account in the calculations, the planes are exactly equal. Actually, the fact that the habit plane of the bcc→fcc transformation is the same as the one of the bcc→fcc transformation is logical and comes directly from the criterion. Since in our analysis, the HP is assumed to be an unrotated plane; this plane is the same for both direct and inverse transformations.

There are few experimental studies on the habit planes in pure bcc→fcc transformations. Ohmori *et al.* [34] report that the Widmanstätten austenite $\gamma$ laths are in KS OR with the ferritic $\delta$ matrix and exhibit a well-defined $(\bar{1}10)_\delta$//$(\bar{1}11)_\gamma$ habit plane with the growth direction parallel to $[111]_\delta$//$[110]_\gamma$, which is completely coherent with the first solution found for $(\theta = 1.571, \phi = 2.356)$ . The same OR and the same HP were observed for the bcc→fcc transformation obtained by heating martensitic steels to produce reverse austenite [41][42]. In Cu-Zn brass, Srinivasan and Heworth [38] investigated the habit planes by Laue diffraction and reported two possible different HPs indexed in the parent $\alpha$ bcc phase: $(2, 11, 12)_\alpha$ and $(123)_\alpha$ with a large scatter of the results depending on the alloy composition; but interestingly the scatter is not random and actually the HPs are aligned in the pole figure on a segment containing the $<111>_\alpha$ dense direction and located between the two extreme $(2, 11, 12)_\alpha$ and $(123)_\alpha$ planes (Fig. 3 of ref. [38]). Unfortunately, the authors did not precise the corresponding OR without the ambiguities of the parent symmetries, which impedes a direct comparison with our calculations. We can just notice here that the $(-12, 11, 2)_\alpha$ plane is at 7.4° from the calculated $(\bar{1}10)_\alpha$ HP, and the $(\bar{3}21)_\alpha$ HP is at 4.3° from the calculated $(\bar{5}32)_\alpha$ HP, and that the common $<111>_\alpha$ dense direction is in good agreement with the neutral line chosen for the calculations. It could be worth studying more in details the HPs in Cu-Zn alloys to get more precise and statistical experimental results for comparison. It should be also acknowledged that the assumptions taken in our calculations are probably too rudimentary: the atoms in Cu-Zn alloys do not have the same size, which means that the hard-sphere model with a unique size is not appropriate, and the daughter phase is an orthorhombic or monoclinic 9R distorted fcc structure.

## 6. BCC→HCP transformations

Bcc→hcp transformations occur in Ti and Zr alloys; the former are widely used in aerospace, medical and sport industries [43], and the latter for fuel cladding in nuclear reactors [44]. To keep the coherency of the paper, we will continue to name the bcc phase $\alpha$, and the hcp phase $\varepsilon$, although they are not the usual notation for these alloys (they are generally noted β and $\alpha$, respectively). The crystallography of the bcc→hcp transformation in zirconium has been investigated by Burgers in 1934 [17] and his model is considered as an important reference in metallurgy, such as the Bain model [45] for fcc→bcc transformations. Burgers has clearly identified the most important macroscopic characters of a martensitic transformation: the transition is “homogeneous” and “oriented”. At the microscopic scale, the Burgers' model combines a shear parallel to a $\{112\}_\alpha$ plane in a $<111>_\alpha$ direction, with a shuffle and a homogeneous contraction of the lattice. This lattice distortion is obtained by considering an orthorhombic superlattice close to the hcp and bcc lattices; which is used by Bowles and Mackenzie in the PTMC calculations to predict the habit planes [46], and later by other researchers [47][48]. Another approach based on the edge-to-edge matching (E2EM) model has also been proposed by Zhang *et al.* [49]. In the Burger's paper it is possible to figure out what are the atomic displacements of the atoms; however, it is possible to improve the model by introducing the fact that the atoms are hard-spheres and by merging the discontinuous steps into a continuous mechanism, such as was done in the previous sections for the other phase transformations.

### 6.1. Matrix of complete lattice distortion

As for fcc→hcp transformation, since the primitive unit cell of the hcp phase contains two atoms, it is not possible to find a homogenous distortion that transforms a bcc crystal into a hcp crystal, and thus a shuffle is required for half of the atoms in the lattice. By considering Fig. 5 and Fig. 6b, it appears that the vectors **PO** =½ $[111]_\alpha$ and **PN** = $[\bar{1}10]_\alpha$ remain invariant during the transformation, and that only vector **PK** =½ $[11\bar{1}]_\alpha$ is rotated such that the angle (**P0**,**PK**) which is initially 70.5° decreases to 60°. The natural choice of the primitive basis $\mathbf{B}_p^\alpha$ is therefore **x** = **PO** =½ $[111]_\alpha$ , **y** = **PK** =½ $[11\bar{1}]_\alpha$ and **z** = **PN** = $[\bar{1}10]_\alpha$. They are transformed into the vectors **PO** = $[100]_\varepsilon$ , **PK** = $[110]_\varepsilon$ and **PN** = $[001]_\varepsilon$ respectively, as illustrated in Fig. 5 . This means that $\mathbf{B}_p^\alpha = \begin{bmatrix} 1/2 & 1/2 & -1 \\ 1/2 & 1/2 & 1 \\ 1/2 & -1/2 & 0 \end{bmatrix}$ in the basis $\mathbf{B}_0^\alpha$ has for image, in $\mathbf{B}_0^\varepsilon$ the basis of the product phase, the basis $\mathbf{B}_{p\ \varepsilon 0}^{\alpha'} = \begin{bmatrix} 1 & 1 & 0 \\ 0 & 1 & 0 \\ 0 & 0 & 1 \end{bmatrix}$.

This image can be calculated in the reference basis of the parent crystal $\mathbf{B}_0^\alpha$ by using the correspondence matrix $\mathbf{C}_0^{\alpha\to\varepsilon} = [\mathbf{B}_0^\alpha \to \mathbf{B}_0^\varepsilon]$ given in Table 1. It follows that $\mathbf{B}_p^{\alpha'} = \mathbf{C}_0^{\alpha\to\varepsilon}\,\mathbf{B}_{p\ \varepsilon 0}^{\alpha'}$ and then $\mathbf{D}_0^{\alpha\to\varepsilon} = \mathbf{C}_0^{\alpha\to\varepsilon}\mathbf{B}_{p\ \varepsilon 0}^{\alpha'}\left(\mathbf{B}_p^\alpha\right)^{-1}$, which becomes after calculations

$$\mathbf{D}_0^{\alpha\to\varepsilon} = \begin{bmatrix} \frac{1}{16}(14+\sqrt{6}) & \frac{1}{16}(-2+\sqrt{6}) & \frac{1}{8}(2-\sqrt{6}) \\ \frac{1}{16}(-2+\sqrt{6}) & \frac{1}{16}(14+\sqrt{6}) & \frac{1}{8}(2-\sqrt{6}) \\ \frac{1}{8}(3-\sqrt{6}) & \frac{1}{8}(3-\sqrt{6}) & \frac{1}{4}(1+\sqrt{6}) \end{bmatrix} \quad (34)$$

### 6.2. Matrix of continuous transformation

The distortion matrix of bcc→hcp transformation can be calculated with the method used for the fcc→bcc transformation [14]. Let us consider the non-orthogonal frame $\mathbf{B}_p^\alpha$ constituted by the normalized axes **x** = $(1/\sqrt{3})[111]_\alpha$, **y** = $(1/\sqrt{3})[11\bar{1}]_\alpha$, and **z** = $(1/\sqrt{2})[\bar{1}10]_\alpha$ as illustrated in Fig. 5. The $[111]_\alpha$ and $[11\bar{1}]_\alpha$ directions define the $(\bar{1}10)_\alpha$ plane that is transformed into the $(001)_\varepsilon$ plane by the distortion. Now, let us associate the orthonormal basis $\mathbf{B}_s^\alpha$ = ($\mathbf{x}_s$, $\mathbf{y}_s$, $\mathbf{z}_s$) to the basis $\mathbf{B}_p^\alpha$ = (**x**, **y**, **z**) as usually done for the structural tensor. The coordinates of the **x**, **y** and **z** vectors in the basis $\mathbf{B}_s^\alpha$ give the correspondence matrix from $\mathbf{B}_s^\alpha$ to $\mathbf{B}_p^\alpha$, which is function of the angle β:

$$\left[\mathbf{B}_s^\alpha \to \mathbf{B}_p^\alpha(\beta)\right] = \begin{bmatrix} 1 & \cos(\beta) & 0 \\ 0 & \sin(\beta) & 0 \\ 0 & 0 & 1 \end{bmatrix} \quad (35)$$

The distortion matrix can be expressed in the basis $\mathbf{B}_p^\alpha$ by

$$\mathbf{D}_p^{\alpha\to\varepsilon}(\beta) = \left[\mathbf{B}_p^\alpha(70.5°) \to \mathbf{B}_p^\alpha(\beta)\right] = \left[\mathbf{B}_p^\alpha(70.5°) \to \mathbf{B}_s^\alpha\right]\left[\mathbf{B}_s^\alpha \to \mathbf{B}_p^\alpha(\beta)\right] \quad (36)$$

This matrix can be expressed in the reference basis $\mathbf{B}_0^\alpha$ by

$$\mathbf{D}_0^{\alpha\to\varepsilon}(\beta) = \left[\mathbf{B}_0^\alpha \to \mathbf{B}_p^\alpha(70.5°)\right] \mathbf{D}_p^{\alpha\to\varepsilon}(\beta) \left[\mathbf{B}_0^\alpha \to \mathbf{B}_p^\alpha(70.5°)\right]^{-1} \quad (37)$$

which becomes after calculations

$$\mathbf{D}_0^{\alpha\to\varepsilon}(\beta) = \frac{1}{4}\begin{bmatrix} 3 + X + \frac{\sqrt{1-X^2}}{\sqrt{2}} & -1 + X + \frac{\sqrt{1-X^2}}{\sqrt{2}} & 2 - 2X - \sqrt{2}\sqrt{1-X^2} \\ -1 + X + \frac{\sqrt{1-X^2}}{\sqrt{2}} & 3 + X + \frac{\sqrt{1-X^2}}{\sqrt{2}} & 2 - 2X - \sqrt{2}\sqrt{1-X^2} \\ 1 + X - \sqrt{2}\sqrt{1-X^2} & 1 + X - \sqrt{2}\sqrt{1-X^2} & 2 - 2X + 2\sqrt{2}\sqrt{1-X^2} \end{bmatrix} \quad (38)$$

with X = cos(β).

One can check that for the initial state, since β = 70.5° (X=1/3), $\mathbf{D}_0^{\alpha\to\varepsilon}(70.5°)$ is the identity matrix. The complete transformation is obtained for β = 60° (X = 1/2); which leads to the matrix (34).

### 6.3. Schuffle

All the atoms of type N = (*u*, *v*, *w*) located as N in the plane $(\bar{1}10)_\alpha$ in even layers, i.e such that *l = (-u+v)* is even, have a trajectory that follows directly the lattice distortion:

$$\mathbf{PN'} = \mathbf{D}_0^{\alpha\to\varepsilon}(\beta)\,\mathbf{PN} \quad (39)$$

All the other atoms of type M = (*u*, *v*, *w*) located as M in the plane $(\bar{1}10)_\alpha$ in odd layers, i.e. such that *l = (-u+v)* is odd, have a trajectory that does not follow directly the lattice distortion, but that can be deduced of it. They obey the same global movement as in the bcc→hcp distorted lattice except that they describe locally in their cells the same trajectory as for the bcc→fcc transformation $\mathbf{D}_0^{\alpha\to\gamma}(\beta)$, as illustrated in Fig. 5b. Two shuffle directions are possible: one in which M moves towards P, as shown in Fig. 6b, and one in which it moves in the opposite direction towards R. The origin P of the unit cell in which the atom M is located is deduced from M by the translation vector **t** = $[010]_\alpha$. Thus, the two possible shuffles of M, noted S1 and S-1, are given by

<u>Shuffle S1:</u> $\mathbf{PM'} = \mathbf{D}_0^{\alpha\to\varepsilon}(\beta)\,(\mathbf{PM}-\mathbf{t}) + \mathbf{D}_0^{\alpha\to\gamma}(\beta)\,\mathbf{t}$ (40)

<u>Shuffle S-1:</u> $\mathbf{PM'} = \mathbf{D}_0^{\alpha\to\varepsilon}(\beta)\,(\mathbf{PM}-\mathbf{t}) - \mathbf{D}_0^{\alpha\to\gamma}(\beta)\,\mathbf{t}$, with **t** = ½ $[010]_\alpha$ (41)

Therefore, contrarily to the bcc→fcc transformation, but as for the fcc→hcp transformation, the bcc→hcp transformation requires a shuffle of half of the atoms in the lattice. The two equivalent shuffles were already noticed by Burgers in his early work [17], even if not analytically expressed as here. They are at the origin of the stacking faults in the hcp laths or plates that were observed by TEM in titanium alloys [43].

### 6.4. Prediction of the habit planes

The matrix (34) gives the images of the directions by the bcc→hcp distortion in the reference basis $\mathbf{B}_0^{\alpha}$. The images of the planes are given by the inverse of its transpose:

$$(\mathbf{D}_0^{\alpha\to\varepsilon})^* = {}^T(\mathbf{D}_0^{\alpha\to\varepsilon})^{-1} = \frac{1}{18}\begin{bmatrix} 15+\sqrt{6} & -3+\sqrt{6} & 6-3\sqrt{6} \\ -3+\sqrt{6} & 15+\sqrt{6} & 6-3\sqrt{6} \\ 6-2\sqrt{6} & 6-2\sqrt{6} & 6(1+\sqrt{6}) \end{bmatrix} \quad (42)$$

The two matrices $\mathbf{D}_0^{\alpha\to\varepsilon}$ and $(\mathbf{D}_0^{\alpha\to\varepsilon})^*$ give the images of the directions and planes in the initial basis $\mathbf{B}_0^{\alpha}$. These images can be calculated in the final base $\mathbf{B}_0^{\varepsilon}$ by using the correspondence matrix $C_0^{\alpha\to\varepsilon}$ given in Table 1. The images of the low-index directions and planes are given in Table 3.

As in previous sections, it is also possible to determine the habit plane as the plane unrotated by the distortion. The rotation amplitude is given as a function of the spherical coordinate angles θ and ϕ in Fig. 8. Here six solutions are found. They can be grouped into two triplets due to the fact that the solutions at θ > π/2 are equivalent to those obtained at θ ≤ π/2 because of the centrosymmetric equivalence of lines (θ, ϕ) ≡ (π-θ, ϕ+π). Thus, there are three unequivalent solutions which are $(\bar{1}\bar{1}2)_\alpha$, $(\bar{1}10)_\alpha$ and $(-0.6123, -0.6123, 0.5)_\alpha$ planes. Contrarily to all the phase transformations described in the previous sections, the plane of the third solution does not contain the neutral line. It is at 0.5° of the rational plane $(\bar{5}\bar{5}4)_\alpha$ and it is transformed into $(340)_\varepsilon$ by the distortion matrix. There are a lot of scatter in the habit planes reported in titanium and zirconium alloy, but many studies report the $\{443\}_\alpha$ habit planes (it is recalled again that α is the bcc phase in our study). Although a direct comparison is difficult because of the symmetries of the parent bcc phase, it can be noticed that the calculated $(\bar{5}\bar{5}4)_\alpha$ plane is at 1.5° from the $(\bar{4}\bar{4}3)_\alpha$ habit plane. However, such good matching could be a coincidence. In a recent paper, Qiu *et al.* [50] deeply investigated by TEM martensitic plates in a Ti-Cr alloy analysis and indexed unambiguously all their diffraction patterns according a particular Burgers OR chosen among the equivalent ones. The habit plane "of type M plate" they investigated, written with our choice of Burgers OR, is $(\bar{4}\bar{4}5)_\alpha$ // $(130)_\varepsilon$. The calculated $(\bar{5}\bar{5}4)_\alpha$ // $(340)_\varepsilon$ habit plane is at the 12° far from their reported $(\bar{4}\bar{4}5)_\alpha$ // $(130)_\varepsilon$ habit plane. This agreement is qualitatively good because both calculated and reported HPs are perpendicular to the hcp basal plane, but not quantitatively sufficient for the moment to compete with the PTMC [47][48] or E2EM [49][50] models. The quantitative discrepancy could be due to the fact that in real alloys the hard-sphere approximation is not sufficiently accurate anymore because of the difference of size of the Ti atoms in the bcc and hcp phases, and because the transformation occurs in an alloy and not in a mono-atomic phase.

The continuous analytical expressions of the angular distortive matrices of bcc→fcc and bcc→hcp transformations have been represented with VPython. Simulation "movies" of the distortion of a bcc cube constituted by 8x8x8 unit cells transformed into fcc or hcp structure are given in Supplementary Materials. The initial, intermediate and final states are represented in blue, yellow and red colors, respectively, in Fig. 9a and b.

## 7. Algebra of transformations in the fcc-hcp-bcc system

The calculations for the different transformations in the fcc-hcp-bcc system can seem tedious; however, the similarities of the mathematical treatments make possible a simple encoding with only triplets of simple numbers. The first number in the triplets is taken among (+1,-1,0) and represents the variation of the angle β (between 60° and 70.5°), i.e. it is +1 for Δβ = 70.5°-60° and -1 for Δβ = 60°-70.5°. For example, the fcc-bcc transformation is written with the +1 and the inverse transformation with the -1. The second and third values in the triplets represent the displacement of the atoms that are initially in the position $l = 1$ and $l = 2$ (atoms M and N, respectively). For an initial fcc phase, the number 1 means that the atom moves such that the difference between the final and initial positions is the vector -1/6.$[211]_\gamma$, and 2 means vector -1/3.$[211]_\gamma$. The letter S is added to specify that a shuffle of the atom M ($l$ = 1) is required for the transformation with the hcp phase. Since the shuffle can be obtained by two equivalent ways, there are two equivalent codes. This notation permits to synthetically write the transformations in the fcc-hcp-bcc system, as illustrated in Fig. 10. The fcc→bcc transformation is (1,1,2), fcc→hcp is (0,S0,2) or equivalently (0,S2,2), bcc→hcp is (-1,S1,0) or equivalently (-1,S-1,0). The neutral operation is (0,0,0). The operation of the inverse transformation is coded by multiplying all the values in the triplet by -1, for example bcc→fcc is (-1,-1,-2). The operations can be combined with the additive rule, for example fcc→bcc followed by bcc→hcp is (1,1,2) + (-1,S1,0) = (0,S2,2) = fcc→hcp. However, the equivalency of the shuffles in the hcp phase can generate a problem. For example, if we choose fcc→hcp = (0,S2,2) and hcp→fcc = (0,S0,-2) and compose them, then we would obtain fcc→fcc = (0,S2,0) instead of (0,0,0), and the (0,S2,0) operation is impossible because it would suppose that the atom in M moves to the position N while the atom already in N remains at the same position, i.e. there would be two atoms in the same position. Therefore, one must create two distinct sets of operations:

- Set 1: (fcc→bcc) = (1,1,2), (fcc→hcp) = (0,S0,2) , (bcc→hcp) = (-1,S-1,0), with their inverse, and their neutral elements (fcc→fcc), (bcc→bcc) and (hcp→hcp), all equal to (0,0,0).
- Set 2: (fcc→bcc) = (1,1,2), (fcc→hcp) = (0,S2,2) , (bcc→hcp) = (-1,S1,0), with their inverse, and their neutral elements (fcc→fcc), (bcc→bcc) and (hcp→hcp), all equal to (0,0,0).

One may think that the algebraic structure formed by each set with the additive composition law is a group, but that is not exactly the case. Some operations can't be composed, for example (fcc→bcc)(fcc→bcc) has no meaning. Actually, it can be noticed that the operations are "arrows" and that two operations can be combined if and only if the finish phase of the first arrow is the starting phase of the second arrow; for example (fcc→bcc) (bcc→hcp) = (fcc→hcp). Moreover, each arrow has actually two neutral operations, one at the left, and one at the right; for example (fcc→bcc) has for left neutral element (fcc$\rightarrow$fcc) and for right neutral element (bcc$\rightarrow$bcc) , both encoded by (0,0,0). The algebraic structures of sets 1 and 2 are groupoids. More information on groupoids can be found in ref. [18]. Here, the (fcc$\rightarrow$fcc) twinning operation is not included, but its encoding would be (0, 2, 4). Fig. 10 is a mathematical updated version of the old Burgers scheme (scheme p573 of ref. [17]). The groupoid of simple transformations on the fcc-hcp-bcc system could constitute the core a larger groupoid of all the transformations in the fcc-hcp-bcc system including the twinning modes in each of the fcc, hcp and bcc phases, the orientational variants due to the symmetries (details are given in section 8.6), and their compositions by transformation cycling. This groupoid should be very complex and the most difficult task to build it will be to find a general encoding rule. Indeed, for example, the encoding rule chosen for fcc-fcc multiple $\Sigma 3^n$ twinning [18] is for the moment not composable with the triplet coding presented in this section.

## 8. Discussion

### 8.1. Vocabulary and associated concepts

Before starting the discussion, and in order to avoid any misunderstanding that could be source of controversy, we think important to precise the concept behind the words that we will be used. We will equivalently employ the terms “transformation” and “transition”. The term “transition” is used more generally by physicists, and “transformation” by metallurgists; the former treats the cases of phase change with no or only short-range order rearrangements of the atoms, whereas the latter also includes long range diffusion and therefore precipitation. The term “displacive” was initially attributed to transformation involving the displacement of an atom in its unit cell; and since this effect is inevitably correlated to the distortion of the lattice, both terms “martensitic” and “displacive” can be used equivalently.

The term “reconstructive” needs clarification because it may lead to important confusions, as already noticed by Otsuka and Ren in their review paper on Ti-Ni shape memory alloys [51]. In crystallography, “reconstructive” means that some of the atomic bounds in the parent phase are broken and new bounds are formed in the daughter phase; some symmetry elements of the parent phase are lost and new ones are “reconstructed” in the daughter phase [52]. In metallurgy, the term “reconstructive” is synonymous of “diffusive”, and thus includes long range ordering or precipitation mechanisms; it is often used in opposition to the term “displacive” [9][10]. All the fcc-hcp-bcc transformations treated in the present paper are classified as both “reconstructive” and “displacive” in crystallography, but only as “displacive” in metallurgy. In the rest of the discussion the term “reconstructive” will be used with its crystallographic meaning. In metallurgy, displacive transformations are always associated to formation of surface relief at a polished surface. However this phenomenon is at the origin of one of the most important controversy which has split the community of metallurgists into two groups: the “shearists” and the “diffusionists”, as Zhang and Kelly succinctly called them in their review paper [53]. To briefly summarize, the “shearists” assume that a surface relief can be created only by a displacive mechanism [9]- [12], whereas the “diffusionists” think that in some alloys it is created by diffusion, with the formation of “terraces of growth ledges” with an atomic correspondence at the parent/daughter interface [54][55]. Both groups have developed their own crystallographic tools to predict the orientation relationships and habit planes; i.e. the PTMC for the former, and the E2EM for the latter [53].

The "diffusionists" often use the term "precipitation" in their studies on lath or plate formation in the fcc-hcp-bcc systems. However, the term "precipitation" will be considered here in a very strict meaning, such as it is in aluminum alloys [56]. Some atoms of specie Y that are in solid solution in a matrix constituted of atoms of specie X, diffuse and migrate toward each other due to their chemical driving forces; first, they make small clusters which then grow slowly each time a new atom Y joins the cluster. The atoms Y, associated or not to the atoms X or to other species, form a new crystallographic structure which is in OR with the matrix to minimize the interfacial misfits. Sometimes a reordering of the atoms in the precipitate structure during the precipitation growth occurs due to a size effect [57]. During their growth the precipitates become semi-coherent, and incoherent at micron-scale. Since the atoms come from all around the surrounding cluster, the precipitation mechanism (called "reconstructive" by the metallurgists) is isotropic, and the precipitate shape is only a consequence of the symmetries of the precipitate and matrix phases (see also section 8.2). The precipitated clusters are generally metastable and dissolve during thermal treatments at high temperatures, and the Y species reprecipitate to form new stable phases. It is clear that the approach and equations described in the present paper do not apply to precipitation.

The term "twinning" has very broad meaning that comes from the early Friedel's work on mineralogy; he defines: " *A twin is a complex edifice built up of two or more homogeneous portions of the same crystal species in contact (juxtaposition) and oriented with respect to each other according to well-defined law*" [58](see also ref. [59]). It means that in a polycrystalline material any misorientation found with a frequency higher than it could be expected from a random distribution of isotopic orientations is a "twin". This definition includes annealing twins, mechanical (micro- and macro-) twins, and it also includes the specific misorientations that exist between the variants after a phase transformation, i.e. the transformation twins. However, to our point of view, annealing and mechanical twins are slightly different from transformation twins. In the former case a crystal $\gamma_0$ of phase γ is transformed into another crystal $\gamma_1$ of same phase, whereas in the latter case, a parent crystal $\gamma_1$ of phase γ is transformed into many distinct variants $\alpha_i$ of phase $\alpha$, and the misorientation between two variants $\alpha_i$ can be only understood by considering their parent crystal. In the first versions of PTMC [7]-[10], the lattice invariant shears are mechanical twins (or dislocations) whereas in its advanced versions [11] it includes the "twin" boundaries between pairs of (self-accommodating) variants. In the rest of the discussion, the term "twin" and "twinning" will only apply to mechanical twinning.

### 8.2. Considerations on symmetries

Even if the approach presented in this paper does not apply to precipitation, some concepts on symmetries are more easily introduced by considering the case of precipitation. The orientational symmetries of the shape of a precipitate of phase $\alpha$, embedded in a matrix of phase $\gamma$, are the symmetries common to both matrix and precipitate [60][61]; they are given by the intersection group $\mathbf{H}^\gamma$ between the two point groups $\mathbf{G}^\alpha$ and $\mathbf{G}^\gamma$, taking into account the OR between them. More exactly:

$$\mathbf{H}^\gamma = \mathbf{G}^\gamma \cap \mathbf{C}_0^{\gamma\to\alpha}\, \mathbf{G}^\alpha\, (\mathbf{C}_0^{\gamma\to\alpha})^{-1} \qquad (43)$$

where $\mathbf{C}_0^{\gamma\to\alpha}$ is the correspondence matrix introduced in equation (3)

In displacive transformations, an additional element should be taken into account: all the atoms move coherently in “the same direction” at the speed of sound if the transformation is not diffusion-controlled. This oriented lattice transformation is given by a “shear” vector if one assumes that martensitic transformation are shear transformation, or, in the present approach, they are given by the direction of the rotation of the vector **z** involved in calculation of the distortion matrix. Therefore, even if the symmetries common to the parent and daughter crystals are always given by $\mathbf{H}^{\gamma}$ in equation (43), as for precipitation, the distortion mechanism introduces an additional symmetry breaking, and the shape of the daughter martensite is a subgroup of $\mathbf{H}^{\gamma}$. For example, in fcc→hcp transformations $\mathbf{H}^{\gamma}$ contains 12 symmetries elements. In the case of precipitation, since one of them is the three-fold axis normal to the common dense plane $(\bar{1}11)_{\gamma}$ // $(001)_{\varepsilon}$, the hcp precipitates have a triangular shape. In the case of martensitic transformation, only one of the three equivalent $<112>_{\gamma}$ vector is the “shear” vector at the origin of the transformation, or equivalently in our approach, only one of the three equivalent **z** = **PN** should be chosen to be transformed into the $[001]_{\varepsilon}$ axis, and thus $\mathbf{H}^{\gamma}$ is broken by this “choice” into a subgroup $\mathbf{K}^{\gamma}$ containing 4 elements. The hcp variants have a shape of symmetries $\mathbf{K}^{\gamma}$ , and not $\mathbf{H}^{\gamma}$ , and therefore are not triangular anymore, as illustrated in Fig. 11. For the fcc→bcc transformation with the KS OR, $\mathbf{H}^{\gamma}$ only contains the identity and inversion elements, and thus can’t be broken by the “shearing” component. Actually, if martensite is created by cooling, the three “shear” modes can operate successively on layers of thickness around 50 nm to accommodate the distortion strains [62]. However, if the transformation is triggered by tensile straining [63], only one shear mode will be activated so that the distortion strain accommodates the tensile strain.

### 8.3. The main idea in the one-step model

The main idea in our approach is that the atoms move during the transformation as if they roll on each other’s, and it is these movements which generate the lattice distortion. The mechanism operates whatever the deformation modes of the parent matrix in which the daughter martensite forms. A good image of such approach is the solidification of water in a rigid bottle: if the undercooling is sufficiently high, the dilatation induced by the phase change breaks the bottle whatever its constituent material (actually, if the bottle can withstand internal pressures higher than 220 MPa, a new phase of ice denser than water will form). This concept is very basic but has many implications in our model of martensitic transformations. For example, all the distortion matrices that have been calculated are independent of the exact plastic deformation modes of the matrix, contrarily to the PTMC approaches which includes the shear relaxation modes in the core of the theory [7]. As already schematized in ref. [14], we are convinced that the global deformation generated by the arrays of accommodating dislocations is a direct consequence of the lattice distortion and not of the details of the accommodation modes. Of course the exact nature of the dislocations (screw, edge, partial etc.) depends on the structure of the parent phase, but the global orientation gradients do not. An experimental observation that supports this point of view is the fact that in the fcc-bcc system, the continuous features observed in the Electron BackScatter Diffraction (EBSD) pole figures of the variants belonging to the same parent grains are similar whatever the direction of the transformation, or, in other words, whatever the plastic deformation mode of the phase, fcc or bcc, i.e. gliding on the $\{111\}_{\gamma}$ planes for fcc parent phase, or on the $\{110\}_{\alpha}$ and $\{112\}_{\alpha}$ planes for bcc parent phase. This is illustrated in Fig. 12 which shows that the features of the bcc laths generated in a martensitic steels by the fcc→bcc transformation (Fig. 12a) are similar to those of the Widmanstätten laths generated in a brass by the bcc→fcc transformation (Fig. 12b), or to those of the fcc plates generated in duplex steels. These features could be simulated by two continuous rotations with angles varying continuously between 0 and 5.26°; one around $[110]_\gamma$ // $[111]_\alpha$ is called **A**, and the other one around $[\bar{1}11]_\gamma$ // $[\bar{1}10]_\alpha$ is called **B** [13]. The **A** and **B** rotations are supposed to be the trace of the plastic accommodation of the parent phase, and are at the origin of our researches on the mechanisms of martensitic transformations [65][66]. For the fcc→bcc transformation with Pitsch OR, we have shown that the distortion matrix contains these two continuous rotations [13]: it deforms the surrounding parent environment and, when the transformation continues to propagate, it generates variants that are no longer in strict Pitsch OR with the parent grain. Gradients of ORs between the Pitsch, KS and NW ORs appear inside each martensitic variant [67]. The approach developed for the fcc→bcc transformation is for the moment semi-qualitative; a strict mathematical method that allows to extract the rotations from the distortion matrices should be established. The polar decomposition does not seem adequate. For example, in the case of the fcc→hcp transformation, the rotation given by the polar decomposition of the matrix (17) would not be the rotation ($[\bar{1}10]_\gamma$, 19.47°) that one could expect, even by assuming that the distortion results from a simple unidirectional shearing of amplitude $s = 1/2\sqrt{2}$ on the $[\bar{1}12]_\gamma$ direction (Fig. 11d).

### 8.4. The intermediate states and the activation energy

Since the habit planes are calculated with the matrices of complete transformation, the interest of calculating the intermediate states can be questioned. We would like to give two answers. First, the analytical calculations provide an idea of the displacements of all the atoms in the crystal during the transformation; which is satisfying from a theoretical point view and less frustrating than with the PTMC which tells nothing about the atomic displacements. Secondly, they show that some dilatations appear during the transformation in the intermediate states and then come back to zero in the final state. This point is important. Let us consider the case of a transformation triggered by a shear stress. Thermomechanical calculations based on Patel and Cohen's criterion [68] predict that, among all the possible variants, some will be favored because of their negative energy in comparison with the other variants. Our calculations prove that these variants will not be created immediately at very low stress levels because of the energy required to "activate" the intermediate states. This situation is illustrated in Fig. 13 in the case of a 2D twinning derived from Fig. 1. Although the applied stress is a simple shear, the atoms can't be sheared, and a dilatation $\delta$ should occur perpendicularly to the shear plane. The deformation is not simple shear but is angular-distortive, which also means that due to the hard-sphere assumption the crystallographic strain-stress correlation is not linear. In the case where one of the variants favored by the stress field is formed in a rigid surrounding environment, the stress normal to the shear plane will be compressive; and if the deformation is elastically accommodated by the variant, the energy per unit surface of product phase would increase by a factor ½ $E_\alpha \delta^2$, where $E_\alpha$ is the Young modulus of the $\alpha$ phase. Let us continue to illustrate the importance of the activation energy with the case of fcc→fcc twinning treated in chapter 3. It can be checked that $\mathbf{D}_0^{\gamma\to\gamma}$ in equation (12) lets invariant the two directions $[110]_\gamma$ and $[101]_\gamma$ and thus any linear combination of them, and $\mathbf{D}_0^*(\eta)$ lets invariant the $(\bar{1}11)_\gamma$ plane. The crystallographic distortion associated to twinning is an invariant plane strain but not a shear strain. The lattice dilatation is due to the fact that the atom located in M must climb between the atoms O and K (Fig.

14b); it is given by $\delta$ = MH/(√3/3), which also corresponds to the variation of the distance between the planes **g** = $(\bar{1}11)_\gamma$, or more explicitly:

$$\frac{1}{\delta} = \frac{(\mathbf{D}_0^{\alpha\to\varepsilon})^*\mathbf{g}}{\mathbf{g}} \text{ , with } \mathbf{g} = (\bar{1}11)_\gamma \qquad (44)$$

The values of the dilatation $\delta$ have been calculated along the distortion path: they are shown in Fig. 14b. The maximum value is $\delta_{max}$= (√6/4)/(1/√3) = 3√2/4 ≈ 1.06, which means that there is a dilatation of 6% perpendicularly to the twinning plane. In other words twinning can't be obtained at constant volume although the initial and final states have the same volume; there is an energetic barrier between the two states due to the 6% of volume change of the intermediate state. For fcc→bcc transformations the maximum value is $\delta_{max}$ = 1.015 (Fig. 14a); lower than with twinning because of the atoms O and K do not remain in contact during the climb of the atom M.

### 8.5. Does the model apply to "slow" transformations?

The transformations that have many of the characteristics of martensitic transformations but that do not occur suddenly have been subjected to debates and controversies between the "shearists" and the "diffusionists". The present discussion will not bring any new experimental result or theoretical breakthrough and thus should be simply considered as the point of view of the author. As mentioned in section 8.1, precipitation in its strict meaning has no common property with martensitic transformations except the crystallographic correspondence between the parent (matrix) phase and the daughter (precipitated) phase. Therefore, the lath formation in Cu-Zn, Au-Cd, Fe-Pt can't be called precipitation. Most of debates on these alloys concern the habit planes and the relief formed at polished surfaces. Here, we would like to discuss another experimental result. As told in section 8.3, we believe that the orientation gradients that can be observed in the parent phase (the continuous features in the pole figures) are the plastic trace of the transformations mechanism. Since the features formed in the bcc martensitic steels are similar to those formed in slowly cooled Cu-Zn brass, duplex steels, bcc bainitic steels and iron meteorites (in which the cooling rates are few hundreds degrees by million year), we must conclude, as the "shearists", that the mechanism is intrinsically "martensitic" whatever the speed of the transformations. However, as the "diffusionists", we think that the transformation in these alloys is limited by the diffusion. Indeed, the chemical composition of the product laths is slightly different from the parent phase; for example, in a duplex steel, the $\gamma$ laths formed inside the $\delta$ ferritic matrix are depleted in chromium and enriched in nickel [33][34]. Thus, we adopt the current consensual opinion that the transformations in these alloys are "diffusion-limited" displacive. The plausible scenario of phase transformation during cooling can described as follows: (a) in the parent phase at high temperature, since the stable daughter phase is chemically different from the parent phase, and the atoms diffuse and migrate during cooling, driven by their difference of chemical potentials between the parent and daughter phase, (b) they form a region which has the equilibrium chemical composition of the daughter phase but still has the crystallographic structure of the parent phase, and then, suddenly (c) the region is displacively transformed into the daughter phase, as schematically illustrated in Fig. 15. The transformation can't occur progressively while each atom arrives at the interface because a critical size is required for the displacive transformation to go over the energy required to create the interface and the strain field, as for nucleation [69]. Another possible cause of the limited speed of transformation is the kinetics of displacements of the dislocations generated by the lattice distortion. These transformation dislocations generate a back-stress field localized in front of the daughter lath or plate, which makes the transformation more difficult (this is the origin of the hysteresis and difference between $T_0$ and $M_s$ temperatures). At high temperatures, it is possible that the dislocations glide progressively far from the lath tip and then relax the local stress field, which, then, allows the continuation of the phase transformation and lath growth. In steels, the kinetics of displacement of the transformation dislocations depends on the carbon content due to a Cottrell atmosphere around them. It is possible that such an effect occurs in bainitic steels, but that point can't be further developed in this paper. From the arguments exposed in this section, we think that calculations presented in this paper do not apply only to pure martensitic transformations but also applies to diffusion-limited martensitic transformations, such as in brass, duplex steels, iron-nickel meteorites, and to bainitic transformations in steels, and more generally to "kinematics-limited" martensitic transformations. The main difference with the pure "shearist" approach is that we believe that shear is not required and does not take part to the intrinsic mechanism of the transformation.

If one accepts the idea that fcc→hcp transformations and fcc→fcc microtwinning are "kinematics-limited" martensitic transformations, then some experimental observations in these systems can be interpreted with another point of view. For example, it is widely admitted that the hcp plates of fcc twins are formed by a well-organized synchronized collective motion of Shockley partial dislocations created by a pole, but there is no experimental proof of the existence of dislocations spiraling around poles mechanisms (chapter 3). There is a simple way to interpret the observations: instead of considering that the dislocations are the *cause* of the transformations and that the product phase is created by the passage of the dislocations, one can actually consider that the dislocations are the *consequence* of the transformation and that they are created only to accommodate the transformation, as discussed in section 8.3. With this point of view, even in the case of microtwinning induced by shear, Shockley partial dislocations would not be generated directly by the shear stress, but would be the consequence of the creation of the fcc twin which is thermodynamically more stable than the fcc parent crystal because of its new orientation in the stress field.

### 8.6. Reversibiliy: crystallography, morphology and dislocations

In this paper, only the fcc-hcp-bcc system is studied because it is the only system in which the hard-sphere rules can be applied with good approximation. The transformations in this system are crystallographically irreversible. We mean that, if only crystallographic arguments are considered, the reverse transformation should give back more orientations than the initial one. More explicitly, let us imagine that a single crystal $\gamma_0$ is transformed by the $\gamma \rightarrow \alpha$ transformation into $N^\alpha$ equivalent $\alpha_i$ variants $i \in [1, N^\alpha]$, the variants $\alpha_i$ will generate by the $\alpha \rightarrow \gamma$ inverse transformations more orientations than only the initial one $\gamma_0$. This is due to the absence of a group-subgroup relationship in crystallographically reconstructive transformations [52]. Bhattacharya *et al.* showed that the existence of a group-subgroup relation is a necessary condition of reversibility [70]. A mathematical demonstration is simple and was given in ref. [61]: The number $N^\alpha$ of orientational variants $\alpha_i$ formed by the transformation $\gamma \rightarrow \alpha$ is $N^\alpha = |\mathbf{G}^\gamma| / |\mathbf{H}^\gamma|$ where $|\mathbf{H}^\gamma|$ is the order of the intersection group; it is given by equation (43). The number $N^\gamma$ of orientational variants $\gamma_j$ formed by the

inverse transformation $\alpha \rightarrow \gamma$ is $N^{\gamma} = |\mathbf{G}^{\alpha}| / |\mathbf{H}^{\alpha}|$. Since $\mathbf{H}^{\gamma}$ and $\mathbf{H}^{\alpha}$ are isomorph, their orders (number of elements) are equal, i.e. $|\mathbf{H}^{\alpha}| = |\mathbf{H}^{\gamma}|$. In the case of a group-subgroup relation, $\mathbf{G}^{\alpha} = \mathbf{H}^{\alpha} \leq \mathbf{G}^{\gamma}$, and then $N^{\gamma} = 1$, which means that only the orientation of the initial parent crystal can be generated by the inverse transformation. Since this condition is not satisfied in the fcc-hcp-bcc system, the transformations should not be reversible. That is why the first papers reporting an important shape memory effect in the Fe-Mn-Si steels [71] were a real surprise, as mentioned in the end note of paper [72]. The reason is not yet fully clarified, but is probably linked to the particular configuration of transformation dislocations. In the case of fcc$\rightarrow$hcp transformation, the accommodation is obtained by the creation of arrays of Shockley partial dislocations that all glide on parallel $\{111\}_{\gamma}$ planes, contrarily to fcc$\rightarrow$bcc transformations which imposes the creation of two sets of dislocations at the origin of the orientation gradients (disclinations) and continuous rotations **A** and **B** (section 8.3). If the dislocations could have glided far from the daughter hcp plates and been stored in the retained austenitic matrix, it is plausible that the same dislocations could move backward to induce the reverse transformation and then generate the same fcc orientation of the parent crystal. If the dislocations can't move and remain close to the plates, the fcc-hcp transformation could continue in the plastic zones containing these dislocations, which creates gradients of orientations that are inherited back during the inverse transformation. Such phenomenon of inheritance of internal gradients of orientations induced by martensitic transformation cycles was used recently by Omori *et al.* to promote abnormal grain growth and elaborate shape memory materials with millimetric grains [40].

There is another factor that can favor the reversibility in crystallographically reconstructive transformations; it is the morphological reversibility. Indeed, we have assumed in ref. [14] and in the present paper that the habit plane is a plane unrotated by the distortion. This means that, if for the $\gamma \rightarrow \alpha$ transformation between a parent crystal $\gamma_i$ and one of its variant $\alpha_j$ the habit plane is $(h_i k_i l_i)_{\gamma}$ // $(h_j k_j l_j)_{\alpha}$ , then the habit plane for the reverse $\alpha \rightarrow \gamma$ transformation between the crystal $\alpha_j$ and its variant $\gamma_i$ should be the same plane $(h_j k_j l_j)_{\alpha}$ // $(h_i k_i l_i)_{\gamma}$. In other words, the reverse transformation does not require the creation of new habit planes if the transformation is obtained between the same parent/daughter crystals. That morphological effect, in addition to the storage of transformation dislocations, could possibly explain the partial reversibility of the transformations in the fcc-hcp-bcc system. A synthetic table of the equivalent distortion matrices, orientations matrices and habit planes of the variants is given in Table 4 for direct and inverse transformations.

## 9. Conclusions

This paper is a generalization of our previous paper [14] which was dedicated to fcc$\rightarrow$bcc martensitic transformation. It gives for the first time the analytical expressions of the atom displacements and lattice distortions during the fcc$\rightarrow$fcc twinning and during fcc$\rightarrow$hcp, bcc$\rightarrow$fcc and bcc$\rightarrow$hcp martensitic transformations. The resulting equations are summarized in Table 5. They are calculated without any fit of free parameters. The main ideas are:

- The atoms are considered as hard-spheres.

- The distortions can't imply a simple shear because it would make the atoms interpenetrate themselves.
- The distortion matrices are calculated from the orientation relationships.
- Shuffle is required for transformations implying the hcp phase because this phase contains two atoms in its Bravais lattice.
- The habit planes are determined numerically on the assumption that they are unrotated by the lattice distortion. They compare quite well with the experimental results reported in literature.
- The continuous intermediate states are calculated. The results show that an activation energy is required, even in the simple case of fcc→fcc twinning produced by a simple shear stress.
- The lattice distortion and atomic shuffle of the transformations in the fcc-hcp-bcc system can be encoded by triplets forming a groupoid algebraic structure.
- The dislocations are the consequence of the lattice distortion, and not the cause. This is the classical point of view for the fcc→bcc martensitic transformation, but we think it is also true for the fcc→hcp transformation and fcc→fcc microtwinning, whereas it is generally believed that the periodic glide of Shockley partial dislocations produced by a pole mechanism is at the origin of the hcp or fcc microtwins.
- The surrounding parent phase accommodates the distortion whatever its deformation modes (glide and twin systems). In the case of fcc→bcc and bcc→fcc transformations, the plastic accommodation is retained and appears under the form of the continuous rotations **A** and **B**.
- A distinction is done between the morphological variants and the orientational variants because the transformation mechanism can induce a symmetry breaking of the intersection group, as discussed in the case of the fcc→hcp transformations.
- By considering literature, we believe that the approach also applies to the diffusion-limited displacive and bainitic transformations.

We think that this model is a good qualitative approach of the martensitic transformations. Of course, it is limited by its basic assumptions and can't treat the transformations in which the atom size changes significantly between the parent and daughter phases, or in alloys constituted by atoms of different sizes. Twinning in bcc or hcp materials have not been treated here because the numerous twinning systems would have made the study more tedious, but the approach seems sufficiently general to treat these cases.

# Acknowledgments

We would like to show our gratitude to Prof. Roland Logé who let us continue working on this theory at LMTM. PX group is also sincerely thanked for their subsidy of the LMTM laboratory.

# TABLES

| | |
|---|---|
| $C_0^{\gamma\to\alpha} = \begin{pmatrix} \frac{1}{3\sqrt{6}} & \frac{1}{18}(12+\sqrt{6}) & \frac{1}{9}(3-\sqrt{6}) \\ \frac{2}{3}-\frac{1}{3\sqrt{6}} & -\frac{1}{3\sqrt{6}} & \frac{1}{9}(3+\sqrt{6}) \\ \frac{1}{9}(3+\sqrt{6}) & \frac{1}{9}(-3+\sqrt{6}) & -\frac{2\sqrt{\frac{2}{3}}}{3} \end{pmatrix}$ | $C_0^{\alpha\to\gamma} = \begin{pmatrix} \frac{1}{2\sqrt{6}} & 1-\frac{1}{2\sqrt{6}} & \frac{1}{2}+\frac{1}{\sqrt{6}} \\ 1+\frac{1}{2\sqrt{6}} & -\frac{1}{2\sqrt{6}} & -\frac{1}{2}+\frac{1}{\sqrt{6}} \\ \frac{1}{2}-\frac{1}{\sqrt{6}} & \frac{1}{2}+\frac{1}{\sqrt{6}} & -\sqrt{\frac{2}{3}} \end{pmatrix}$ |
| $C_0^{\gamma\to\varepsilon} = \begin{pmatrix} \frac{1}{2} & 0 & \frac{2}{3} \\ \frac{1}{2} & -\frac{1}{2} & -\frac{2}{3} \\ 0 & \frac{1}{2} & -\frac{2}{3} \end{pmatrix}$ | $C_0^{\varepsilon\to\gamma} = \begin{pmatrix} \frac{4}{3} & \frac{2}{3} & \frac{2}{3} \\ \frac{2}{3} & -\frac{2}{3} & \frac{4}{3} \\ \frac{1}{2} & -\frac{1}{2} & -\frac{1}{2} \end{pmatrix}$ |
| $C_0^{\alpha\to\varepsilon} = \begin{pmatrix} \frac{1}{2} & \frac{1}{8}(-2+\sqrt{6}) & -1 \\ \frac{1}{2} & \frac{1}{8}(-2+\sqrt{6}) & 1 \\ \frac{1}{2} & \frac{1}{4}(-1-\sqrt{6}) & 0 \end{pmatrix}$ | $C_0^{\varepsilon\to\alpha} = \begin{pmatrix} \frac{1}{9}(6+\sqrt{6}) & \frac{1}{9}(6+\sqrt{6}) & -\frac{2}{9}(-3+\sqrt{6}) \\ \frac{2}{3}\sqrt{\frac{2}{3}} & \frac{2}{3}\sqrt{\frac{2}{3}} & -\frac{4}{3}\sqrt{\frac{2}{3}} \\ -\frac{1}{2} & \frac{1}{2} & 0 \end{pmatrix}$ |

**Table 1** Correspondence matrices between the fcc (γ), bcc (α) and hcp (ε) phases for KS, Burgers and NS ORs between them.

| **Images of directions** | |
|---|---|
| 3 <100>$_\alpha$ | 1 <100>$_\gamma$ 2 <110>$_\gamma$ |
| 6 <110>$_\alpha$ | 4 <211>$_\gamma$ 2 <100>$_\gamma$ |
| 4 <111>$_\alpha$ | 4 <110>$_\gamma$ |
| 12 <112>$_\alpha$ | 8 <123>$_\gamma$ 4 <210>$_\gamma$ |
| **Images of planes** | |
| 3 {100}$_\alpha$ | 1 {100}$_\gamma$ 2 {110}$_\gamma$ |
| 6 {110}$_\alpha$ | 4 {111}$_\gamma$ 2 {100}$_\gamma$ |
| 4 {111}$_\alpha$ | 4 {210}$_\gamma$ |
| 12 {112}$_\alpha$ | 8 {113}$_\gamma$ 4 {110}$_\gamma$ |

**Table 2** Images of low-index directions and planes by the bcc→fcc transformation with inverse KS OR.

| **Images of directions** | |
|---|---|
| 3 $\langle 100\rangle_{\alpha}$ | 2$\langle 211\rangle_{\varepsilon}$ 1 $\langle 100\rangle_{\varepsilon}$ |
| 6 $\langle 110\rangle_{\alpha}$ | 2 $\langle 2\text{-}11\rangle_{\varepsilon}$ 2 $\langle 230\rangle_{\varepsilon}$ 1 $\langle 210\rangle_{\varepsilon}$ 1 $\langle 001\rangle_{\varepsilon}$ |
| 4 $\langle 111\rangle_{\alpha}$ | 2 $\langle 100\rangle_{\varepsilon}$ 2 $\langle 101\rangle_{\varepsilon}$ |
| **Images of planes** | |
| 3 $\{100\}_{\alpha}$ | 2 $\{102\}_{\varepsilon}$ 1 $\{120\}_{\varepsilon}$ |
| 6 $\{110\}_{\alpha}$ | 2 $\{101\}_{\varepsilon}$ 2 $\{210\}_{\varepsilon}$ 1$\{100\}_{\varepsilon}$ 1$\{001\}_{\varepsilon}$ |
| 4 $\{111\}_{\alpha}$ | 2 $\{1\text{-}24\}_{\varepsilon}$ 1 $\{3\text{-}20\}_{\varepsilon}$ 1 $\{120\}_{\varepsilon}$ |

**Table 3** Images of low-index directions and planes by the bcc→hcp transformation with Burgers OR.

| Transformation | $\gamma\to\alpha$ | $\alpha\to\gamma$ |
|---|---|---|
| Point group of the parent phase | $\mathbf{G}^{\gamma}$ | $\mathbf{G}^{\alpha}$ |
| Correspondance matrix (Orientation Relationship) | $\mathbf{C}_0^{\gamma\to\alpha}$ | $\mathbf{C}_0^{\alpha\to\gamma} = (\mathbf{C}_0^{\gamma\to\alpha})^{-1}$ |
| Intersection group | $\mathbf{H}^{\gamma} = \mathbf{G}^{\gamma} \cap \mathbf{C}_0^{\gamma\to\alpha}\, \mathbf{G}^{\alpha}\, (\mathbf{C}_0^{\gamma\to\alpha})^{-1}$ | $\mathbf{H}^{\alpha} = \mathbf{G}^{\alpha} \cap \mathbf{C}_0^{\alpha\to\gamma}\, \mathbf{G}^{\gamma}\, (\mathbf{C}_0^{\alpha\to\gamma})^{-1} \equiv \mathbf{H}^{\gamma}$ |
| Distorsion matrix | $\mathbf{D}_0^{\gamma\to\alpha}$ | $\mathbf{D}_0^{\alpha\to\gamma} = \mathbf{C}_0^{\alpha\to\gamma}(\mathbf{D}_0^{\gamma\to\alpha})^{-1}\,\mathbf{C}_0^{\gamma\to\alpha}$ |
| Habit plane (unrotated plane) | $(h_i k_i l_i)_{\gamma}$ // $(h_j k_j l_j)_{\alpha}$ | |
| Orientational variants | $\alpha_i = \mathbf{g}_i^{\gamma}\, \mathbf{H}^{\gamma}\, \mathbf{C}_0^{\gamma\to\alpha}, \quad \mathbf{g}_i^{\gamma} \in \mathbf{G}^{\gamma}$ | $\gamma_i = \mathbf{g}_i^{\alpha}\, \mathbf{H}^{\alpha}\, \mathbf{C}_0^{\alpha\to\gamma}, \quad \mathbf{g}_i^{\alpha} \in \mathbf{G}^{\alpha}$ |
| Number of orientational variants | $N^{\alpha} = \lvert\mathbf{G}^{\gamma}\rvert / \lvert\mathbf{H}^{\gamma}\rvert$ | $N^{\gamma} = \lvert\mathbf{G}^{\alpha}\rvert / \lvert\mathbf{H}^{\alpha}\rvert$ |
| Morphological variants (variants of transformation) | $\alpha_i = \mathbf{g}_i^{\gamma}\, \mathbf{K}^{\gamma}\, \mathbf{D}_0^{\gamma\to\alpha}, \quad \mathbf{g}_i^{\gamma} \in \mathbf{G}^{\gamma}$ | $\gamma_i = \mathbf{g}_i^{\alpha}\, \mathbf{K}^{\alpha}\, \mathbf{D}_0^{\alpha\to\gamma}, \quad \mathbf{g}_i^{\alpha} \in \mathbf{G}^{\alpha}$ |
| Number of morphological variants | $N^{\alpha} = \lvert\mathbf{G}^{\gamma}\rvert / \lvert\mathbf{K}^{\gamma}\rvert$ | $N^{\gamma} = \lvert\mathbf{G}^{\alpha}\rvert / \lvert\mathbf{K}^{\alpha}\rvert$ |

**Table 4** Comparison of the crystallographic properties of the direct and inverse transformations. The elements $\mathbf{g}_i^{\gamma}$ are matrices of the point group $\mathbf{G}^{\gamma}$; they should not be confused with elements of the reciprocal lattice (planes) also noted by the letter **g** in the paper. Different distortion matrices can lead to the same orientation of the variants, but to different orientations of the shapes, as it is the case for fcc→hcp transformation (section 8.1), that is why, for such transformations, the orientational and morphological variants should be distinguished. $\mathbf{K}^{\gamma}$ is a subgroup of the intersection group $\mathbf{H}^{\gamma}$ , $\mathbf{K}^{\gamma} \leq \mathbf{H}^{\gamma}$ . For fcc→hcp transformations $\mathbf{K}^{\gamma} < \mathbf{H}^{\gamma}$ . For fcc-bcc transformations $\mathbf{K}^{\gamma} = \mathbf{H}^{\gamma}$.

| **Transformation** | **OR** | **Complete distortion matrix** | **Continuous distortion matrix** | **Shuffle** | **Predicted HP** | **Observed HP** | **Difference** |
|---|---|---|---|---|---|---|---|
| FCC → BCC | KS | Equ. 31 of ref. [14] | Equ. 32 of ref. [14] | No | $(\bar{1}11)_\gamma$<br>$(\bar{2}25)_\gamma$ | $(\bar{1}11)_\gamma$<br>$(\bar{2}25)_\gamma$ | 0°<br>0.5° |
| FCC → FCC | Twin | Equ. (8) | Equ. (12) | No | $(\bar{1}11)_\gamma$ | $(\bar{1}11)_\gamma$ | 0° |
| FCC → HCP | NS | Equ. (17) | Equ. (19) | Equ. (21) + (22) | $(\bar{1}11)_\gamma$ | $(\bar{1}11)_\gamma$ | 0° |
| BCC → FCC | KS | Equ. (25)+(25) | Equ. (30) | No | $(\bar{1}10)_\alpha$<br>$(\bar{5}32)_\alpha$ | $(\overline{12},11,2)_\alpha$<br>$(\bar{3}21)_\alpha$ | 7.4°<br>4.3° |
| BCC → HCP | Burgers | Equ. (34) | Equ. (38) | Equ. (40) + (41) | $(\bar{1}10)_\alpha$<br>$(\bar{1}\bar{1}2)_\alpha$<br>$(\bar{5}\bar{5}4)_\alpha$ | -<br>-<br>$(\bar{4}\bar{4}3)_\alpha$<br>$(\bar{4}\bar{4}5)_\alpha$ | -<br>-<br>1.5°<br>12° |

**Table 5** Summary of the equations obtained in the paper. The details of the KS, NS and Burgers orientation relationships are given in the equations (5).

# FIGURES

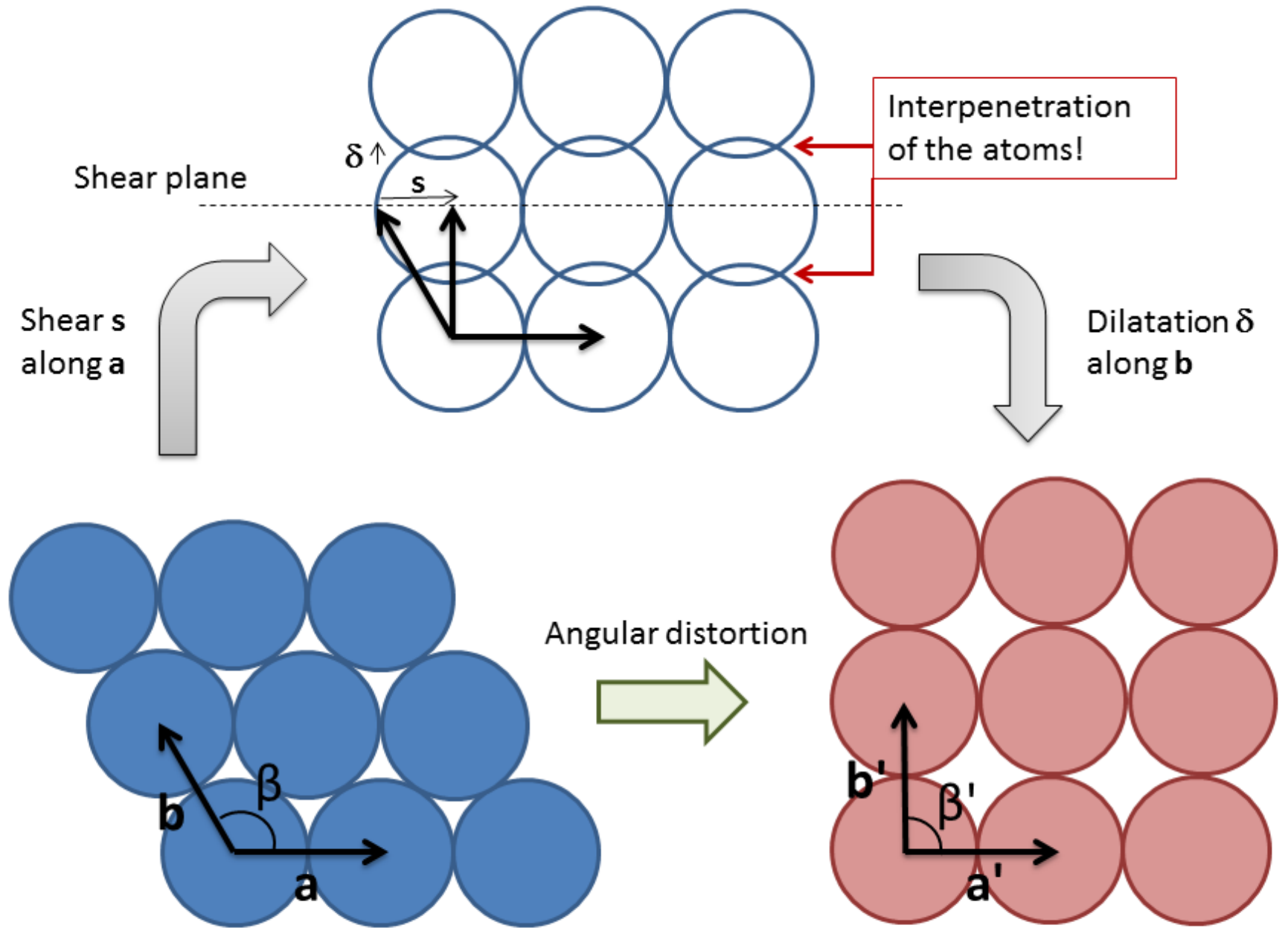

Fig. 1. *2D schematic view of the incompatibility induced by a simple shear deformation in the case of the hard-sphere assumption.*

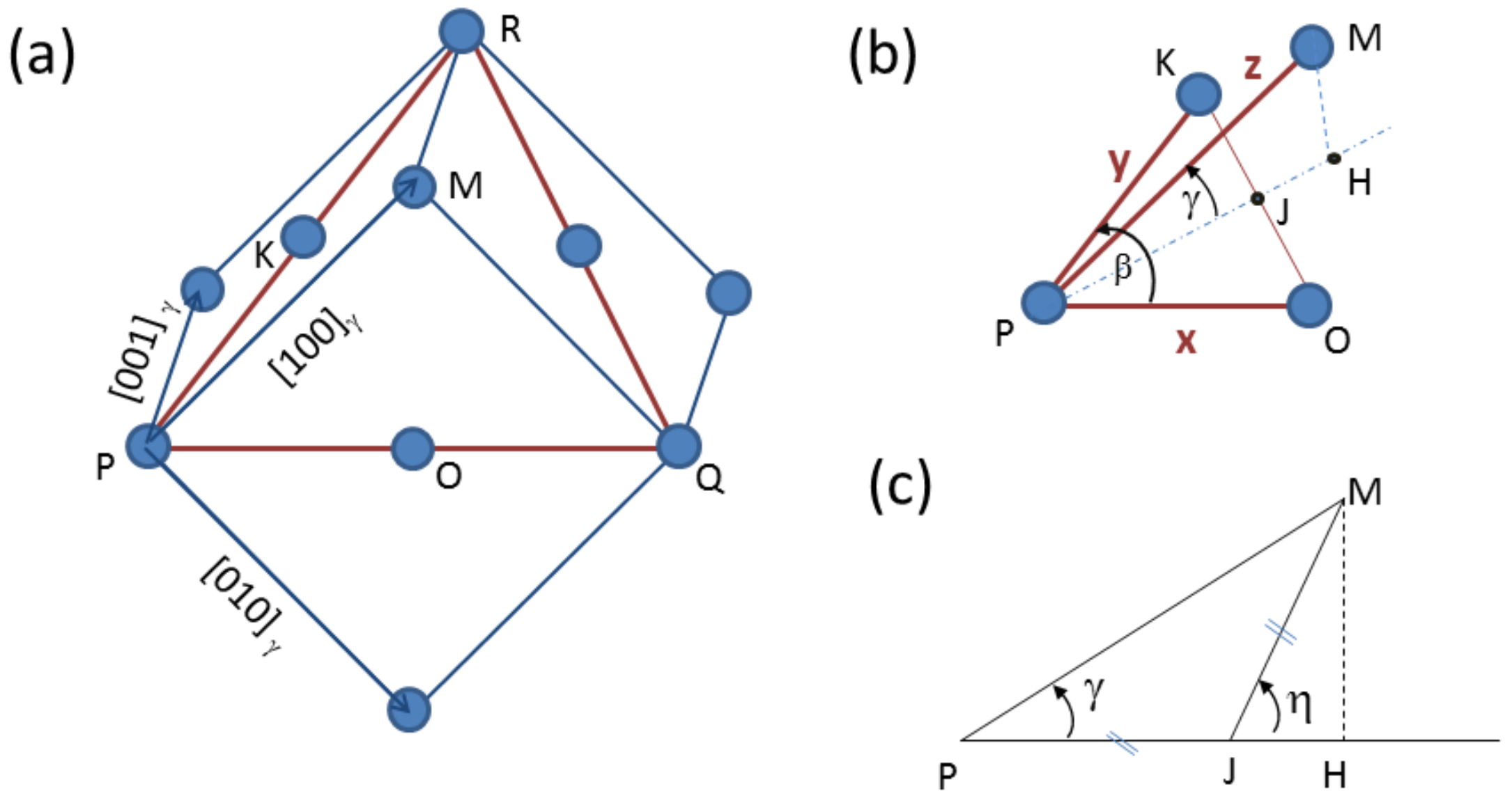

Fig. 2. *Fcc→fcc twinning on the* $(\bar{1}11)_\gamma$ *plane. The plane* $(\bar{1}11)_\gamma$ *is marked by the POK triangle, before twinning, with* ***P0*** = *½* $[110]_\gamma$ *,* ***PK*** = *½* $[101]_\gamma$ *,* ***OK*** = *½* $[0\bar{1}1]_\gamma$ *and* ***PM*** = $[100]_\gamma$ *and* ***JM*** = *¼* $[2\bar{1}\bar{1}]_\gamma$*. The triangle POK is unchanged by twinning (β=60°), contrarily to fcc→bcc transformation. The atom is M, initially such that* ***PM*** = $[100]_\gamma$ *, moves and climbs between the atoms in O and K and, after twinning, in its final position, M is located such that the tetrahedron POKM is regular. During the displacement of point M, the angle η=2γ varies from η =Arcos(1/3)=70.5° to η = - Arcos(1/3)=180°-70.5°.*

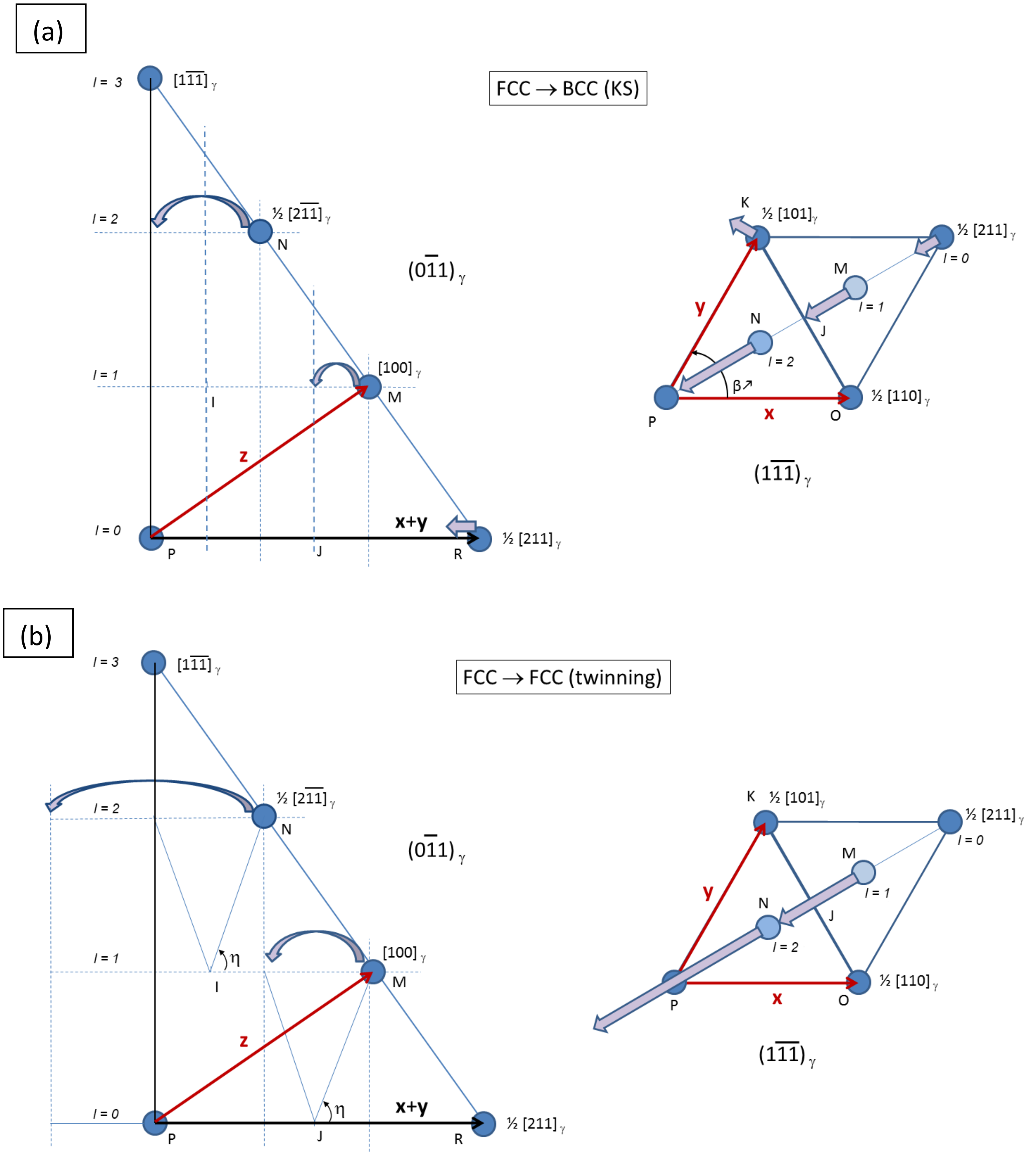

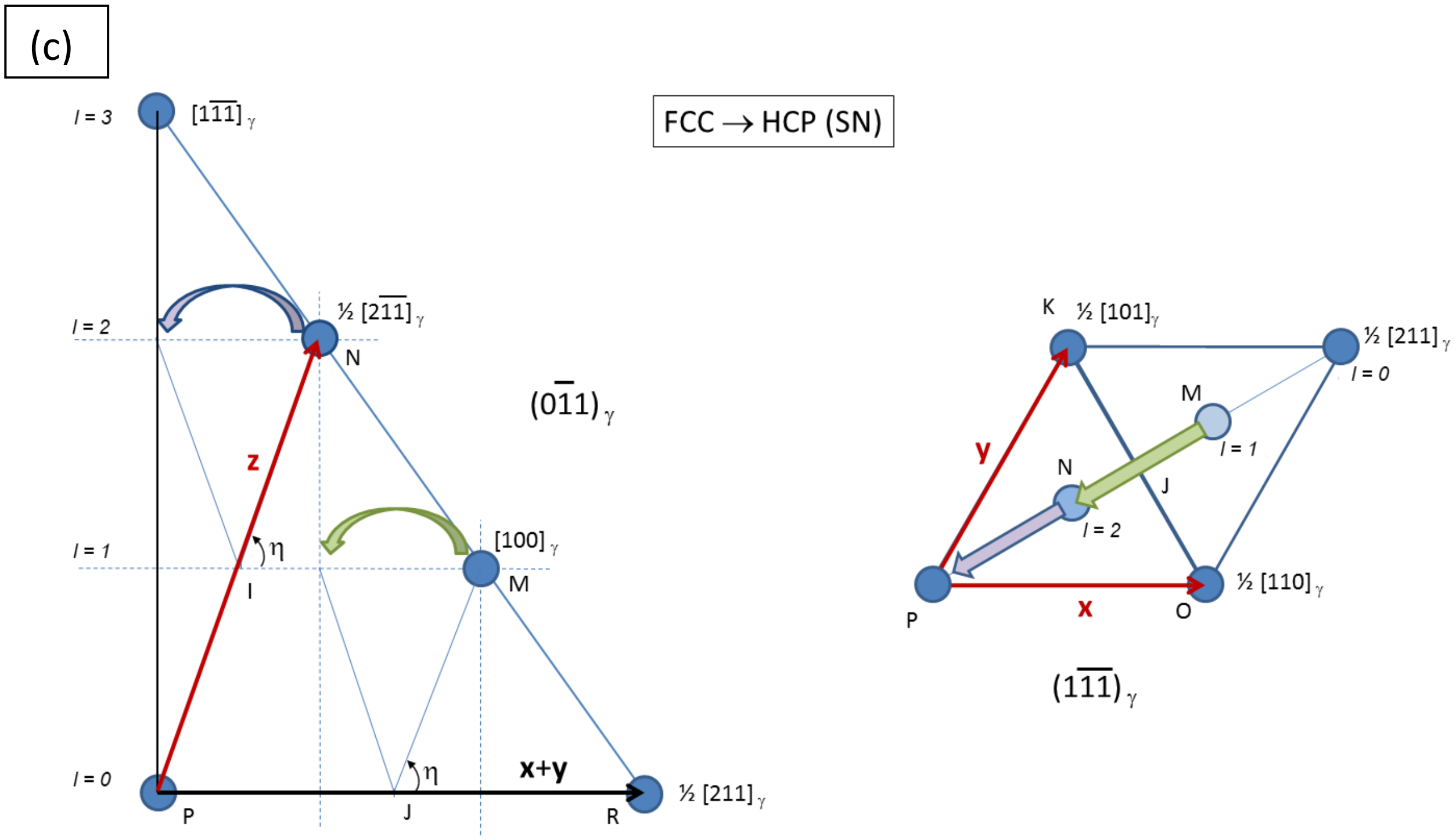

Fig. 3. *Views on the* $(0\bar{1}1)_\gamma$ *and* $(1\bar{1}\bar{1})_\gamma$ *plane of the transformation of a fcc crystal into (a) bcc , (b) mechanically twinned fcc, or (c) hcp crystal, with KS, twin and SN OR, respectively. The purple arrows correspond to the atomic displacements that follow the lattice distortion and the green arrow in (c) represents the shuffle of the atom M. In (c), M moves in the position previously occupied by N, while N moves to the position above the atom P at level l=2. Another possibility of shuffle is that M stays at his position.*

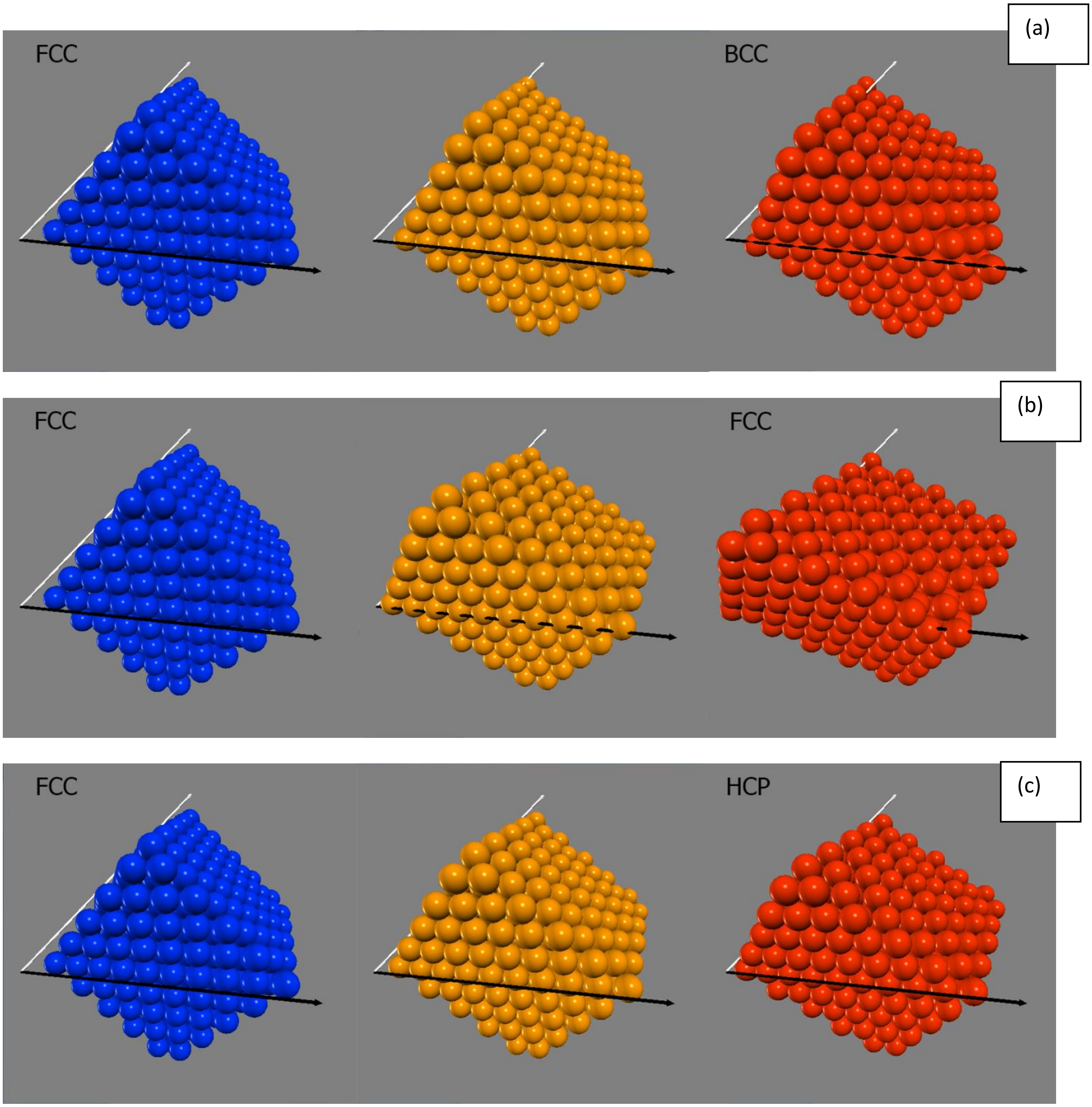

Fig. 4. *3D representations of (a) fcc→bcc martensitic transformation, (b) fcc→fcc mechanical twinning, and (c) fcc→hcp martensitic transformation. In blue, the initial parent fcc crystal with a cube shape with* $\{100\}_\gamma$ *facets. In red, the resulting transformed daughter crystals. In yellow, the intermediate states at medium distortion angle.The black arrow represents the invariant neutral line* $[110]_\gamma$*, and the white arrow the* $[101]_\gamma$ *direction (also invariant for the fcc→hcp and fcc→fcc transformations).*

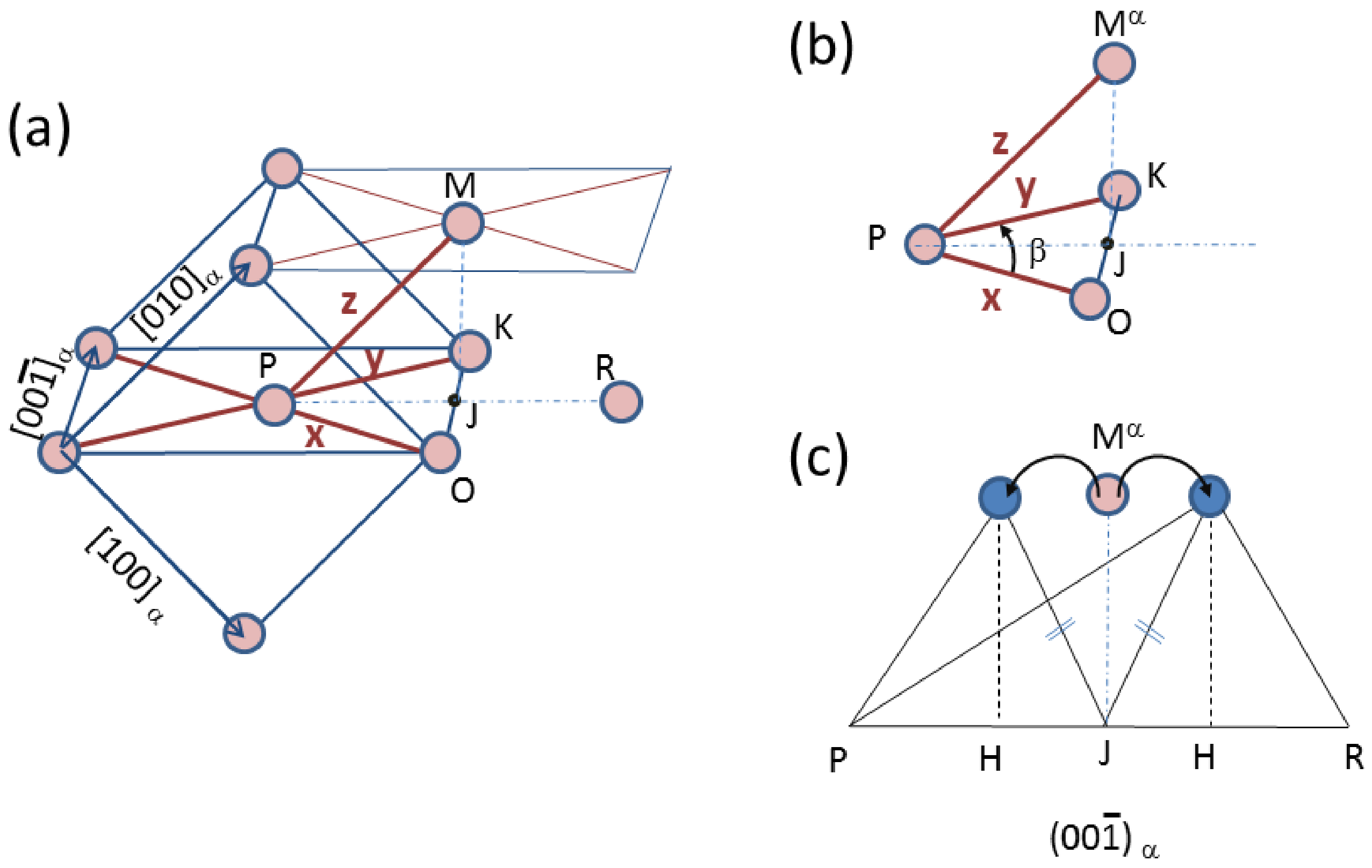

Fig. 5. *Bcc→hcp transformation. (a) The triangle POK corresponds to the same triangle as used for the fcc-bcc transformation. The directions PO and PK are the close-packed directions* ***PO*** *= ½ [111]*$_\alpha$ *and* ***PK*** *= ½ [11$\bar{1}$]*$_\alpha$*. (b) As for fcc→bcc transformation the angle β between these two directions changes from β=60° to β=70.5°. The atom M is such that* ***PM*** *= [010]*$_\alpha$*. The projection of M on the plane POK= ($\bar{1}$10)*$_\alpha$ *is J such that in the bcc structure* ***JM*** *= ½ [$\bar{1}$10]*$_\alpha$*. (c) During the fcc→bcc transformation the atom of the fcc phase initially in position $M^\gamma$ shuffles to the position in $M^\alpha$ of the bcc phase (the distortion of the distance OK is not visible because it is perpendicular to PJ). Two equivalent shuffles in opposite directions are possible.*

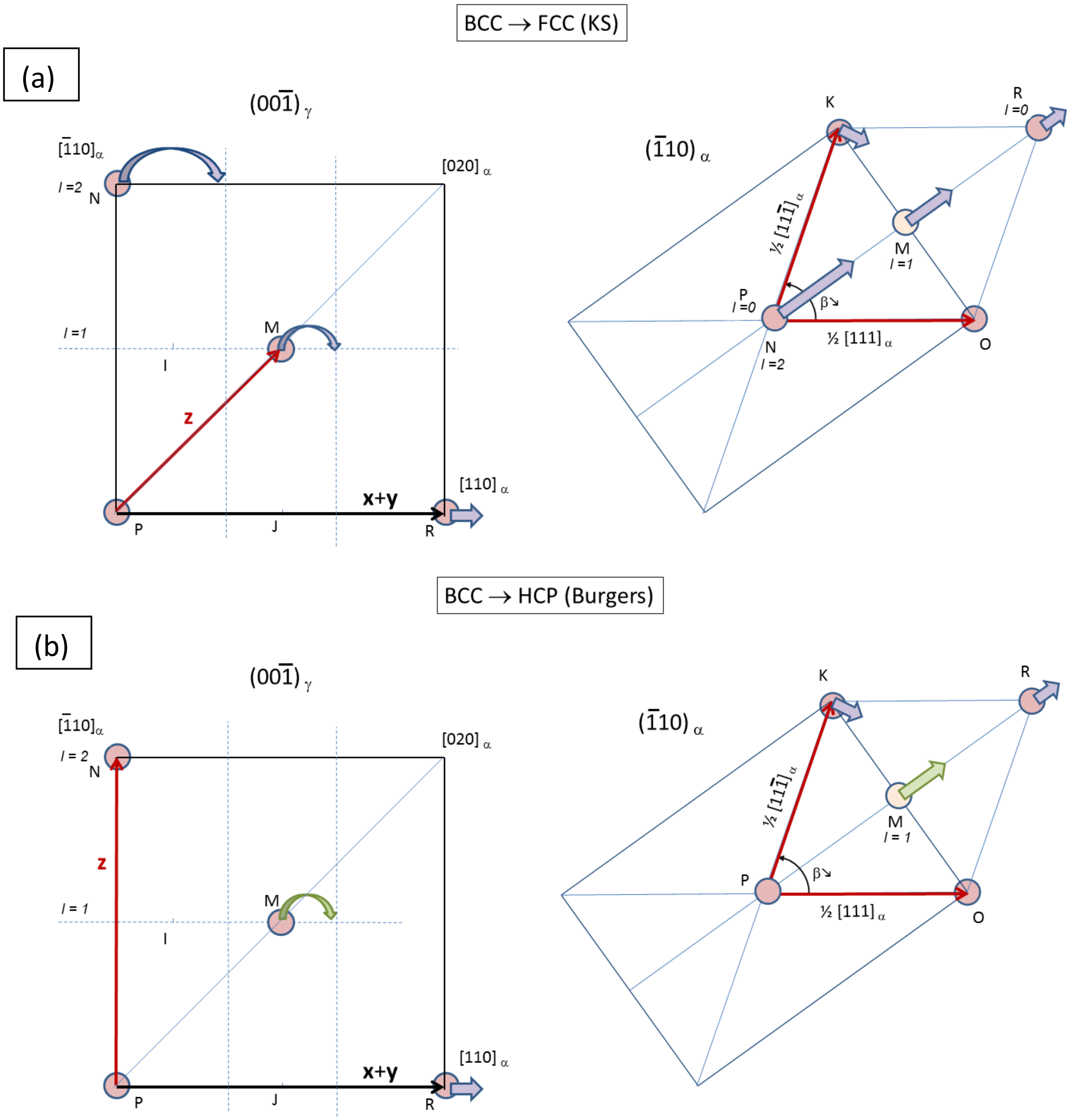

Fig. 6. *Views on the* $(001)_\alpha$ *and* $(\bar{1}10)_\alpha$ *planes of the transformation of a bcc crystal into (a) fcc and (b) hcp crystals, with KS and SN OR, respectively. The purple arrows correspond to the atomic displacements that follow the lattice distortion, and the green arrow in (b) is a shuffle. Only one of the two possible shuffles is represented in (b). The atom in M could also have moved in the opposite direction, as shown in Fig. 5c.*

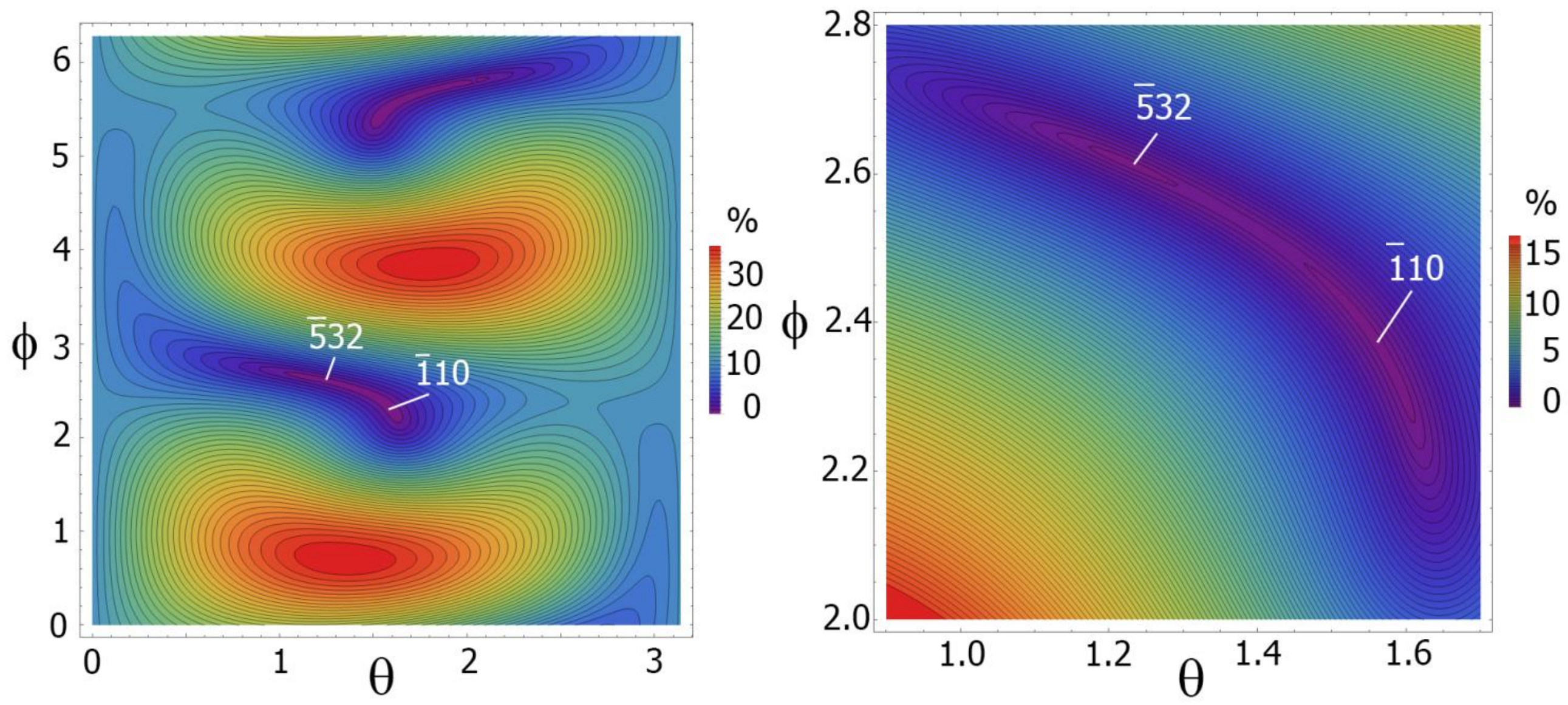

Fig. 7. *Determination of the unrotated planes of the bcc→fcc transformations. (a) Graphical representations with Mathematica of* $\|\boldsymbol{\Delta g}_{\perp}\|$ *given in %. (a) 2D representation according to the spherical coordinates θ and φ, with θ ∈ [0,π] and θ ∈ [0,2π]. (b) Enlargement of the region around the two local minima* $(\bar{1}10)_\alpha$ *and* $(\bar{5}32)_\alpha$.

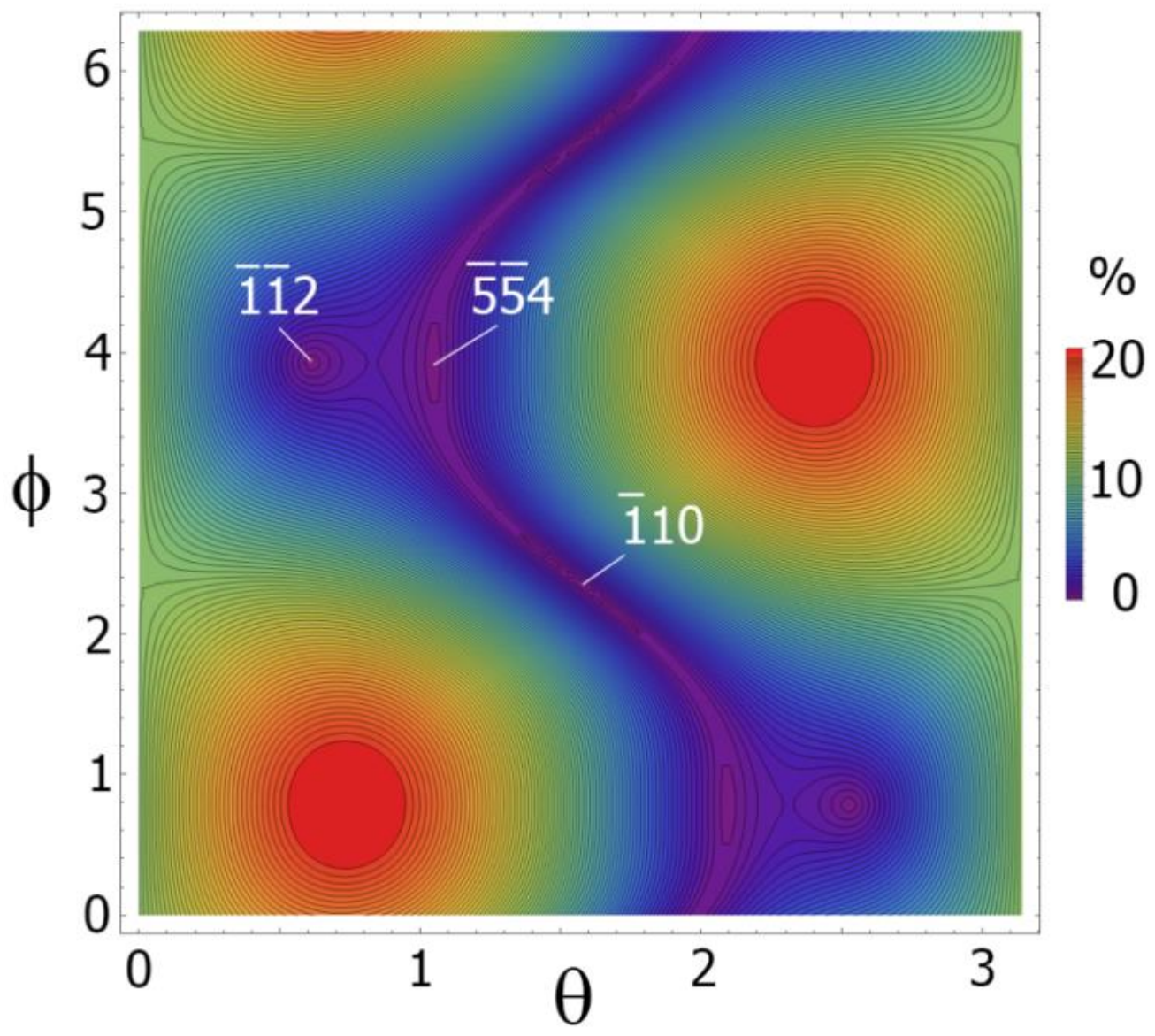

Fig. 8. *Determination of the unrotated planes of the bcc→hcp transformations. (a) Graphical representations with Mathematica of* $\|\boldsymbol{\Delta g}_{\perp}\|$ *given in %. (a) 2D representation according to the spherical coordinates θ and φ, with θ ∈ [0,π] and θ ∈ [0,2π]. The local minima* $(\bar{1}10)_\alpha$ , $(\bar{1}\bar{1}2)_\alpha$ *and* $(\bar{5}\bar{5}4)_\alpha$

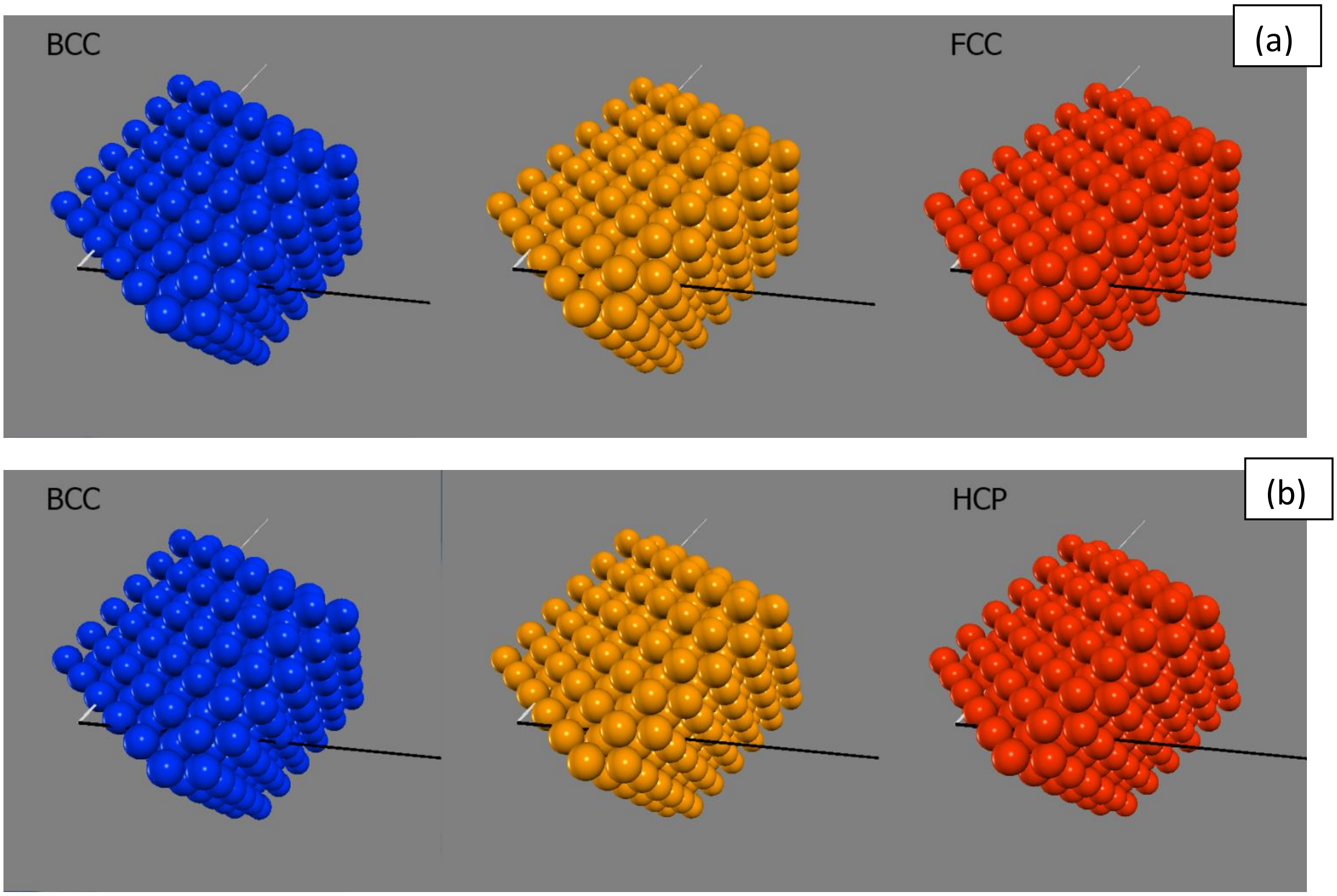

Fig. 9. *3D representations of (a) bcc→fcc and (b) bcc→hcp transformations. In blue, the initial parent bcc crystal with a cube shape with* $\{100\}_\alpha$ *facets. In red, the resulting transformed daughter crystals. In yellow, the intermediate states at medium distortion angle. The black arrow represents the invariant neutral line* $[111]_\alpha$ *, and the white arrow the* $[11\bar{1}]_\alpha$ *direction (rotated by both transformations).*

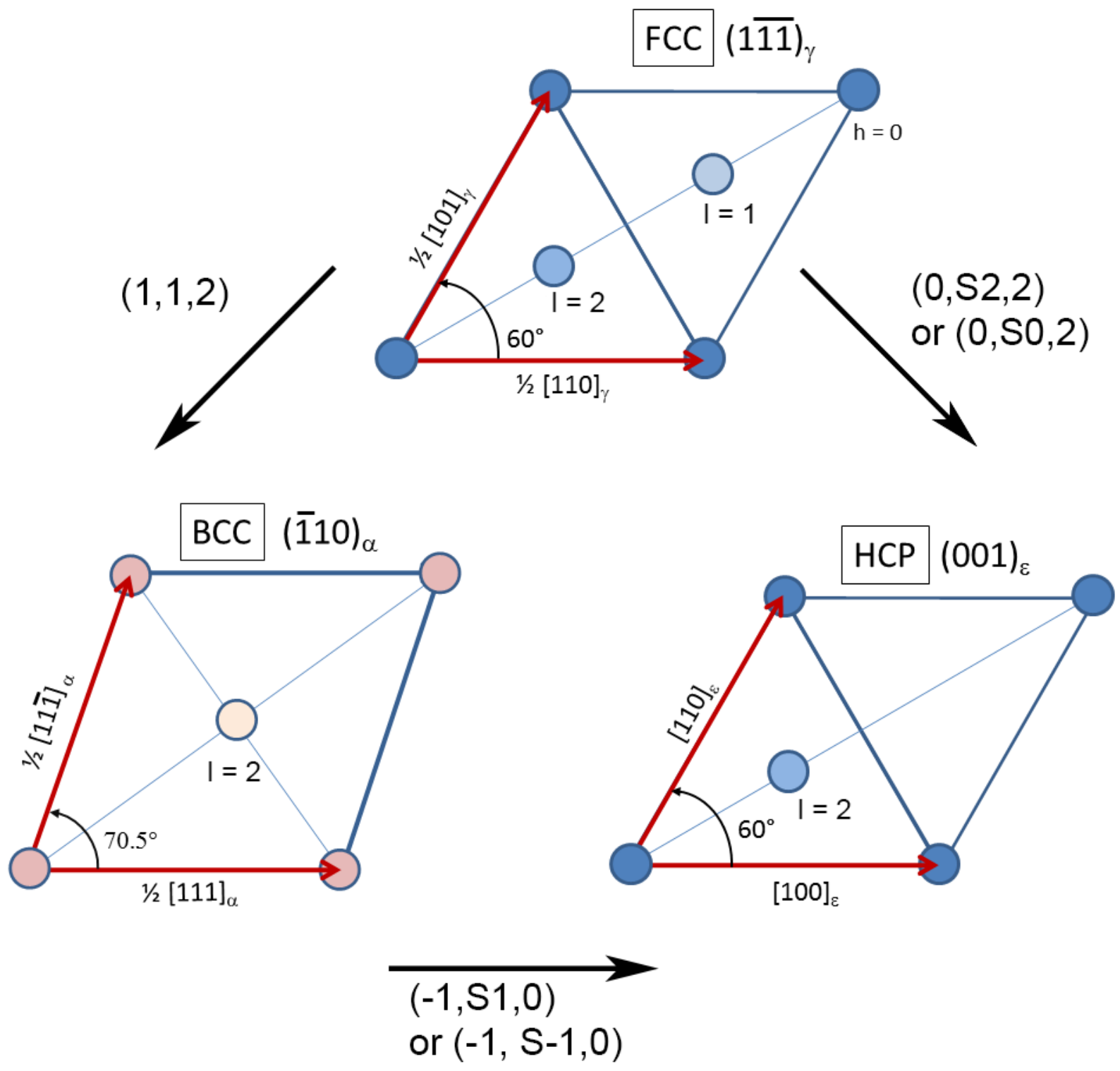

Fig. 10. *Mathematical encoding of the transformations in the fcc-hcp-bcc system. The first number is taken in the list (+1,-1,0) and represents the variation of the angle β (between 60° and 70.5°). The second and third numbers represent the displacement of the atoms that are initially in the position l = 1 and l = 2 (atoms M and N, respectively). The number 1 means that the atom move such that the difference between the final and initial place is the vector -1/6.[211]$_\gamma$ and 2 means -1/3.[211]$_\gamma$. The letter S is added for the atom shuffles.*

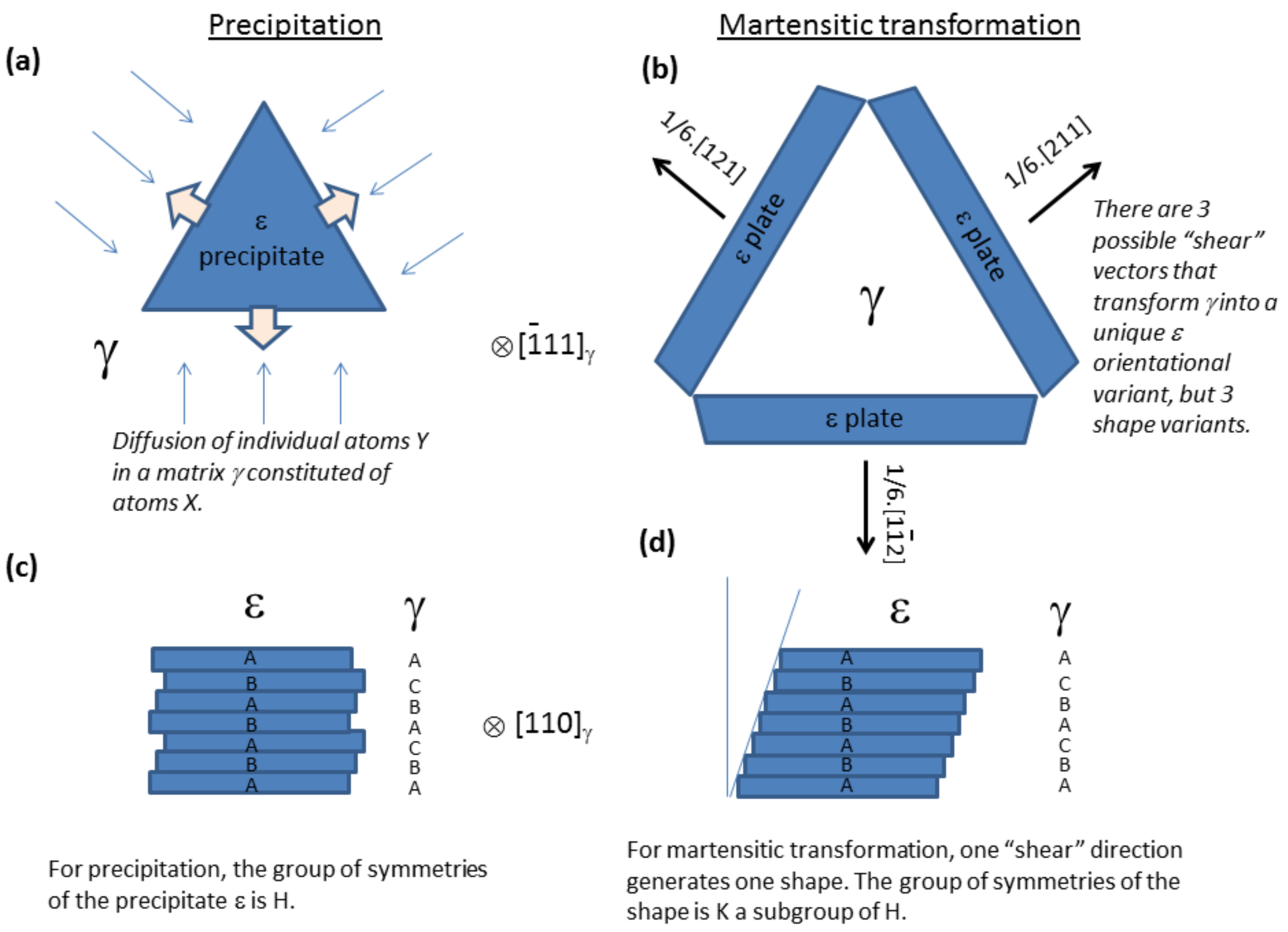

Fig. 11. *Difference of fcc→hcp mechanism between (a,c) precipitation and (b,d) martensitic transformation. The schematic representations are oriented edge-on along* $[\bar{1}11]_\gamma$ *in (a,b), and on the side along* $[110]_\gamma$ *in (b,d). The point group of the shape of the martensitic variants (K) is a subgroup of the point group of the shape of the precipitates (H).*

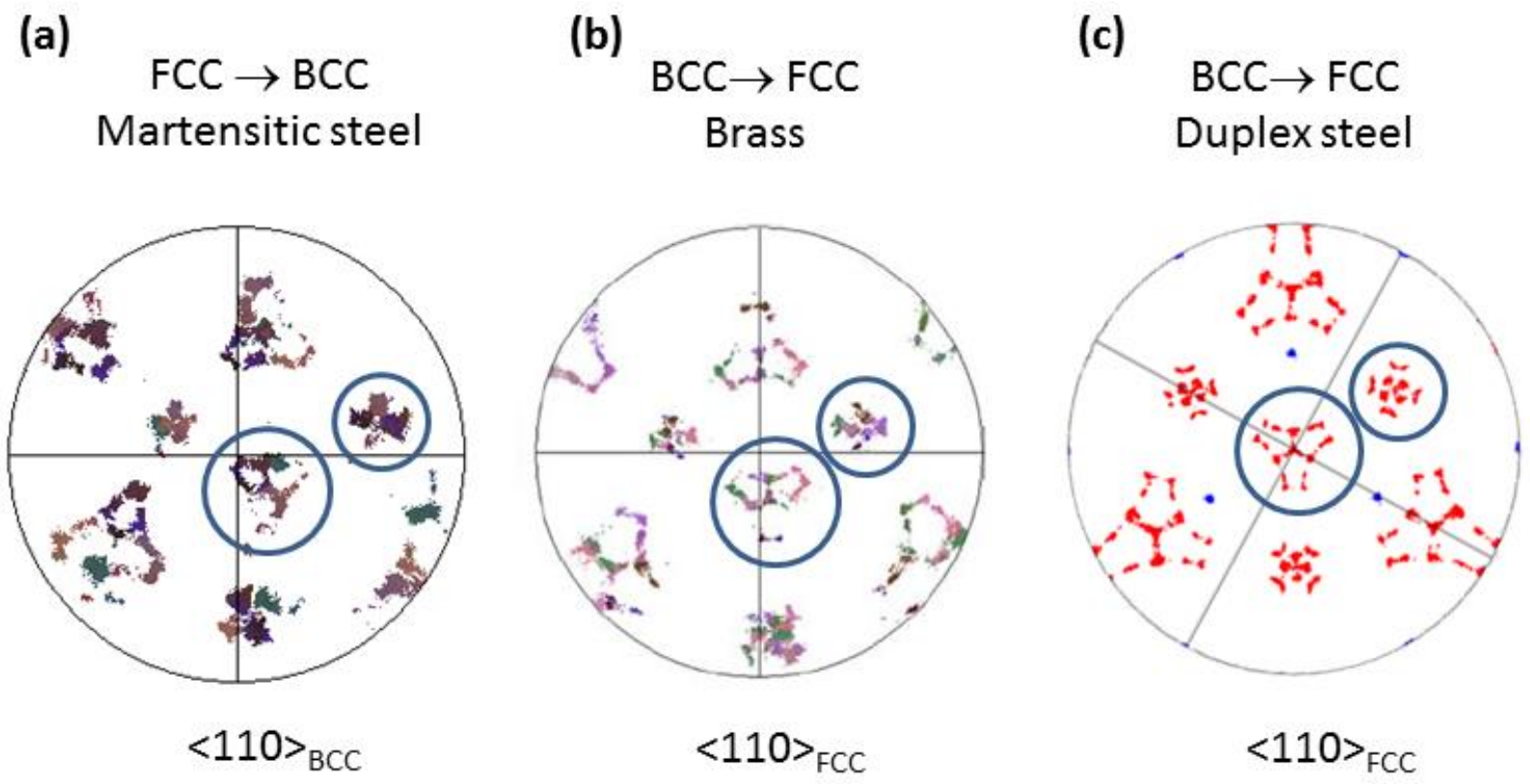

Fig. 12. *Similarities of the continuous features observed in the EBSD pole figures of (a) the <110>$_{bcc}$ directions formed by the martensitic laths in a parent fcc grain of a martensitic steel (EM10, thermally treated), (b) the <110>$_{fcc}$ directions formed by the laths of a Widmanstätten colony in a parent bcc grain of a brass alloy (from [39]), and (c) the <110>$_{fcc}$ directions formed by the martensitic austenite laths in a parent $\delta$ bcc grain, in a duplex steel (from [64]). The three-fold "flower" and four-fold "cross" are identified by the large and medium size circles.*

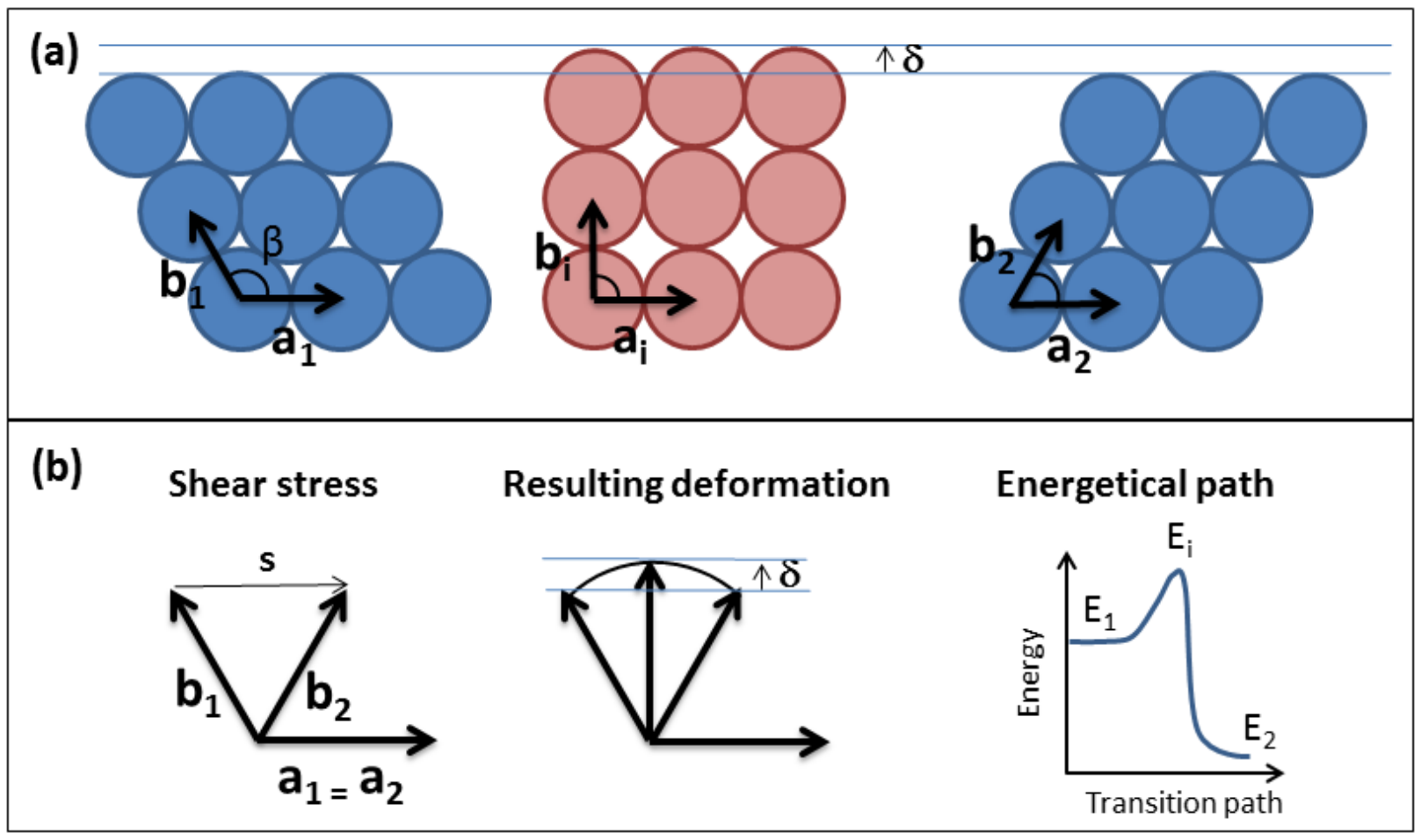

Fig. 13. *Schematic representation of the intermediate states and their importance in the energetical calculations. (a) The ($\boldsymbol{a}_1$,$\boldsymbol{b}_1$) basic vectors of the lattice 1 are transformed by twinning into the ($\boldsymbol{a}_2$,$\boldsymbol{b}_2$) basic vector of the lattice 2 by the application of a shear stress* **s**. *(b) Due to the hard-sphere packing the resulting deformation is not a simple shear strain but an angular distortion, and a slight dilatation component $\delta$ naturally appears during the transformation; it is maximum in the intermediate state in red.*

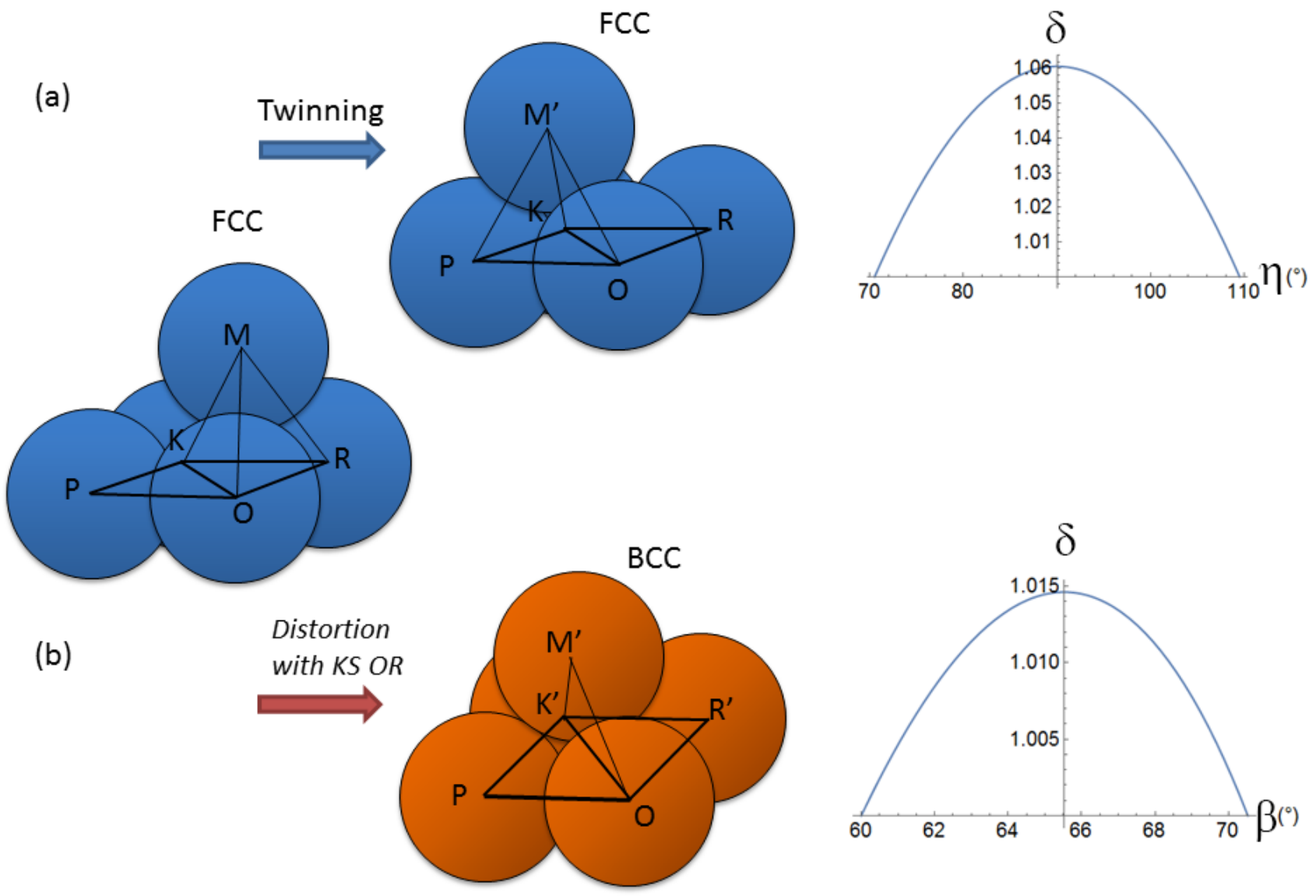

Fig. 14. *Comparison between (a) martensitic transformation and (b) twinning, with a 3D representation of the hard-sphere atoms. The letters P, O, K, M are at the centres of the atoms, as in ref. [13]. The curves at the right side represent the variation of the spacing of the* $(\bar{1}11)_\gamma$ *plane during the transformation, i.e. when the angular order parameter changes from 60° to 70.5° for martensite, and from 70.5° to 109.5° for twinning.*

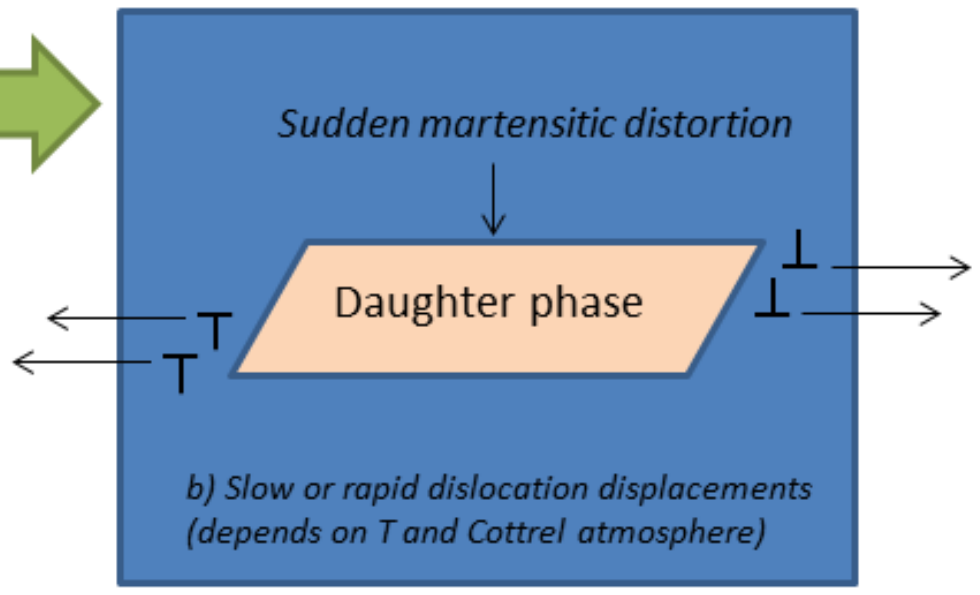

Fig. 15. *Schematic representation of a diffusion-limited displacive transformation. First, the chemical composition of the daughter phase is obtained by a slow process of atomic diffusion, but the crystallographic structure is still the parent phase one, and then, when the critical size is reached, the transformation suddenly (and displacively) occurs. The distortion introduces dislocations in the surrounding parent matrix. The kinetics of the displacements of these dislocations created at the tip of the plate can also influence the kinetics of the transformation.*